\documentclass[twocolumn, superscriptaddress]{revtex4-1} % twocolumn
\newcommand{\SingleColFigScale}{0.9} % twocolumn
\newcommand{\DoubleColFigScale}{0.9} % twocolumn
\newcommand{\BigDoubleColFigScale}{0.9} % twocolumn

\usepackage{graphicx}
\usepackage[usenames,dvipsnames]{xcolor}
\usepackage{amsmath}
\usepackage{amssymb}
\usepackage{mathtools}
\usepackage{bm}
\usepackage{siunitx}
\usepackage{txfonts}
\usepackage[pdfusetitle]{hyperref}
\definecolor{link}{rgb}{0.07, 0.07, 0.80}
\hypersetup{
  colorlinks=true,
  linkcolor=link,
  citecolor=link,
  urlcolor=link
}
\begin{document}

%% \preprint{APS/123-QED}
%% 
%% TITLE
\title{Modeling and analyzing a photo-driven molecular motor system: \\Ratchet dynamics and non-linear optical spectra}
%% AUTHORS
\author{Tatsushi Ikeda}
\email{ikeda.tatsushi.37u@kyoto-u.jp}
\affiliation{Department of Chemistry, Graduate School of Science, Kyoto University, Kyoto 606-8502, Japan}
\author{Arend G.~Dijkstra}
\email{A.G.Dijkstra@leeds.ac.uk}
\affiliation{ School of Chemistry and School of Physics and Astronomy, University of Leeds, Leeds LS2 9JT, United Kingdom}
\author{Yoshitaka Tanimura}
\email{tanimura.yoshitaka.5w@kyoto-u.jp}
\affiliation{Department of Chemistry, Graduate School of Science, Kyoto University, Kyoto 606-8502, Japan}

%% DATE
\date{\today}

\begin{abstract}
  A light-driven molecular motor system is investigated using a multi-state Brownian ratchet model described by a single effective coordinate with multiple electronic states in a dissipative environment.
  The rotational motion of the motor system is investigated on the basis of wavepacket dynamics.
  A current determined from the interplay between a fast photochemical isomerization (photoisomerization) process triggered by pulses and a slow thermal isomerization (thermalization) process arising from an overdamped environment is numerically evaluated.
  For this purpose, we employ the multi-state low-temperature quantum Smoluchowski equations that allow us to simulate the fast quantum electronic dynamics in the overdamped environment, where conventional approaches, such as the Zusman equation approach, fail to apply due to the positivity problem.
  We analyze the motor efficiency by numerically integrating the equations of motion for a rotator system driven by repeatedly impulsive excitations.
  When the timescales of the pulse repetition, photoisomerization, and thermalization processes are separated, the average rotational speed of the motor is determined by the timescale of thermalization.
  In this regime, the average rotational current can be described by a simple equation derived from a rate equation for the thermalization process.
  When laser pulses are applied repeatedly and the timescales of the photoisomerization and pulse repetition are close, the details of the photoisomerization process become important to analyze the entire rotational process.
  We examine the possibility of observing the photoisomerization and the thermalization processes associated with stationary rotating dynamics of the motor system by spectroscopic means, e.g.~pump-probe, transient absorption, and two-dimensional electronic spectroscopy techniques.
\end{abstract}

\pacs{Valid PACS appear here}
\keywords{
  photo-driven molecular rotary motor,
  open quantum dynamics,
  non-adiabatic transition,
  nonlinear optical response,
  low-temperature quantum Smoluchowski equations
}
\maketitle

\section{INTRODUCTION}
\label{sec:introduction}

Light-driven molecular motors are nano-technological machines capable of continuous unidirectional rotation under optical driving \cite{koumura1999nature, feringa2002apa, browne2006nn, feringa2011book}.
Typical photo-driven molecular motor systems involve several processes to accomplish unidirectional rotation, which include photochemical isomerization (photoisomerization) and thermal isomerization (thermalization).
The fast electronic quantum dynamics in photoisomerization after photoexcitation with a typical time scale in the picosecond range \cite{kwok2003jrs, browne2006nn} and slow thermalization in an environment with a typical time scale in the nanosecond to hour range \cite{pollard2007afm, klok2008jacs} are the key to describe the function of photo-driven molecular motor systems.

Experimentally, circular dichroism spectroscopy has been performed to investigate thermalization processes \cite{pollard2007afm, klok2008jacs, klok2009fd}, and various ultrafast nonlinear laser measurements have been carried out to investigate details of photoisomerization processes \cite{conyard2012nc, conyard2014jacs, hall2017jacs}.
These experiments aim to analyze dynamics and yields of each isomerization processes, to understand the roles of their substituents and solvents, and to achieve high performance rotation e.g.~by choosing the substituents and solvents.

Theoretically, many studies have been carried out with the aim of understanding the details of the potential energy surface (PES) structures and non-adiabatic coupling (NAC) profiles to determine the rotational degrees of freedom and to simulate fast photoisomerization processes in molecular motors systems \cite{kazaryan2009jpca, liu2012jacs, amatatsu2013jpca, kazaryan2010jpca, hu2016jcc, kazaryan2011jctc, wang2017jpca, pang2017jpca}.
In contrast, the thermalization process, which is known as a rate-limiting step of molecular motors \cite{pollard2007afm, klok2008jacs, klok2009fd}, has not been well explored.
This is because, most of the thermalization processes of molecular motors can be characterized by simple rate equations, and therefore the detailed dynamics of a single thermalization process is not regarded as an important issue.
However, to discuss the performance of the whole rotation processes, a theoretical description that include both photoisomerization and thermalization is important.
In this paper, we study this entire process using a system-bath model.

A system exhibiting a rectified motion in a periodic potential that utilizes thermal or mechanical fluctuations has been investigated as a ratchet system \cite{smoluchowski1927pms, feynmann2011bookI46, astumian1997science, astumian2002pt}.
The rotational speeds of ratchet systems including biological motor systems such as Myosins and F$_{0}$F$_{1}$-adenosine triphosphate (ATP) synthase, have been intensively investigated from both theoretical and experimental approaches \cite{hanggi2005adp}.
In a molecular motor system, while photoexcitation and thermal fluctuations are almost unbiased, the PESs of the system are asymmetric due to steric hindrance effects of functional groups of the molecule (e.g.~large alkyl and aryl groups).
These asymmetric PESs cause ratchet effects and rectify the current.
Therefore, we can regard a photo-driven molecular motor as a ratchet system \cite{browne2006nn}.
Because the thermalization process should be described using fluctuation and dissipation arising from the environment, here we study the rotary motor system by applying a Brownian heat bath model that has been used for the investigation of the ratchet system \cite{riemann2002pr, hanggi2005adp, hanggi2009rmp, kato2013jpcb}.
%%%
However, it is difficult to adopt the conventional theoretical frameworks which have been used for Brownian ratchet systems to the present motor problem, because the characteristic timescales of processes involved in this system, i.e.~the photoexcitation, photoisomerization, and thermalization processes, are very different, and because the dynamics among multiple electronic states should be described in the framework of quantum mechanics as in many previous theoretical investigations concerning photoinduced dynamics \cite{domcke1997acp, hahn2000jpcb, kuhl2002jcp, balzer2003cpl}.
For example, although Langevin type approaches have commonly been employed to describe various Brownian ratchet systems \cite{riemann2002pr, hanggi2005adp, hanggi2009rmp}, such approaches cannot directly be applied to the investigation of slow relaxation by means of numerical simulation, because convergence of such long-time trajectory sampling is slow.
Moreover, it is difficult to apply such approaches to multi-state systems, because the conventional semi-classical approaches with classical trajectories (e.g.~Fewest switch surface hopping and Ehrenfest methods) give a poor description of non-adiabatic transitions, particularly, when we include a dissipative environment \cite{ikeda2018jctc}.
While quantum Fokker-Planck type approaches are capable of treating such systems, they are computationally more expensive than Langevin type approaches, because their phase space description on the basis of the Wigner distribution function require a huge computational memory.
Moreover, in the case of multi-state systems, a semi-classical treatment of the heat-bath may violate the positivity of the probability distribution for non-adiabatic transition processes:
This phenomenon is known as the positivity problem in open quantum dynamics theories \cite{dumcke1979zpb, pechukas1994prl, frantsuzov1997cpl, tanimura2006jpsj}.

In the strong friction limit of an Ohmic spectral density, Fokker-Planck type approaches are simplified and their phase space distributions reduce to probability distributions in coordinate space.
In this case, we can solve the equations easily with a small computational memory.
In order to avoid the positivity problem, however, we must include non-Markovian quantum low-temperature (QLT) correction terms from the Bose-Einstein (BE) distribution into the theory.
Thus, in this paper, we employ the recently developed multi-state low-temperature quantum Smoluchowski equations (MS-LT-QSE) approach \cite{ikeda2018jctc}, which includes the QLT correction.
This is an overdamped Ohmic limit of the (multi-state) quantum hierarchical Fokker-Planck equations ((MS-)QHFPE) approach \cite{tanimura1991pra, sakurai2011jpca, kato2013jpcb, sakurai2014njp, tanimura2015jcp, ikeda2017jcp}, which is a variant of the hierarchical equations of motion (HEOM) theories \cite{tanimura1989jpsj, tanimura2006jpsj, tanimura2014jcp}.
While general spectral densities can be treated by HEOM theories \cite{kreisbeck2012jpcl, tanimura2012jcp, liu2014jcp, duan2017prb}, here we employ the MS-LT-QSE theory for an Ohmic spectral density to simplify the analysis and to reduce the numerical costs for simulations.

It should be noted that, while roles of conical intersections (CIs) in photoisomerization processes of the photo-driven molecular motor system attract much attention experimentally \cite{conyard2012nc, conyard2014jacs, hall2017jacs} and theoretically \cite{kazaryan2010jpca,kazaryan2011jctc,amatatsu2013jpca,pang2017jpca}, we found the difference between the wavepacket dynamics in internal conversion processes via CI and those via avoided crossing (AC) is minor when the system is subject to a dissipative environment \cite{ikeda2018cp}.
This fact allows us to study both photoisomerization and thermalization using a single coordinate model with AC.

The organization of this paper is as follows.
In Sec.~\ref{sec:theory}, we explain a molecular motor model described by the PESs of the electronic adiabatic states in a strong friction environment.
This model allows us to simulate the wavepacket dynamics for the entire motor rotating process, explicitly.
In Sec.~\ref{sec:results}, we present numerical results to illustrate the relation between the rotational speed of the motor system and the timescales of the photoisomerization/thermalization.
The possibility of observing the photoisomerization and the thermal isomerization is examined by calculating transient absorption, and two-dimensional electronic spectra.
A simple analytical equation to estimate average rotational motor speed is also given.
Section~\ref{sec:conclusion} is devoted to concluding remarks.

\section{THEORY}
\label{sec:theory}
\subsection{Multi-state Brownian ratchet model for a molecular motor system}
\begin{figure*}
  \centering
  \includegraphics[scale=\DoubleColFigScale]{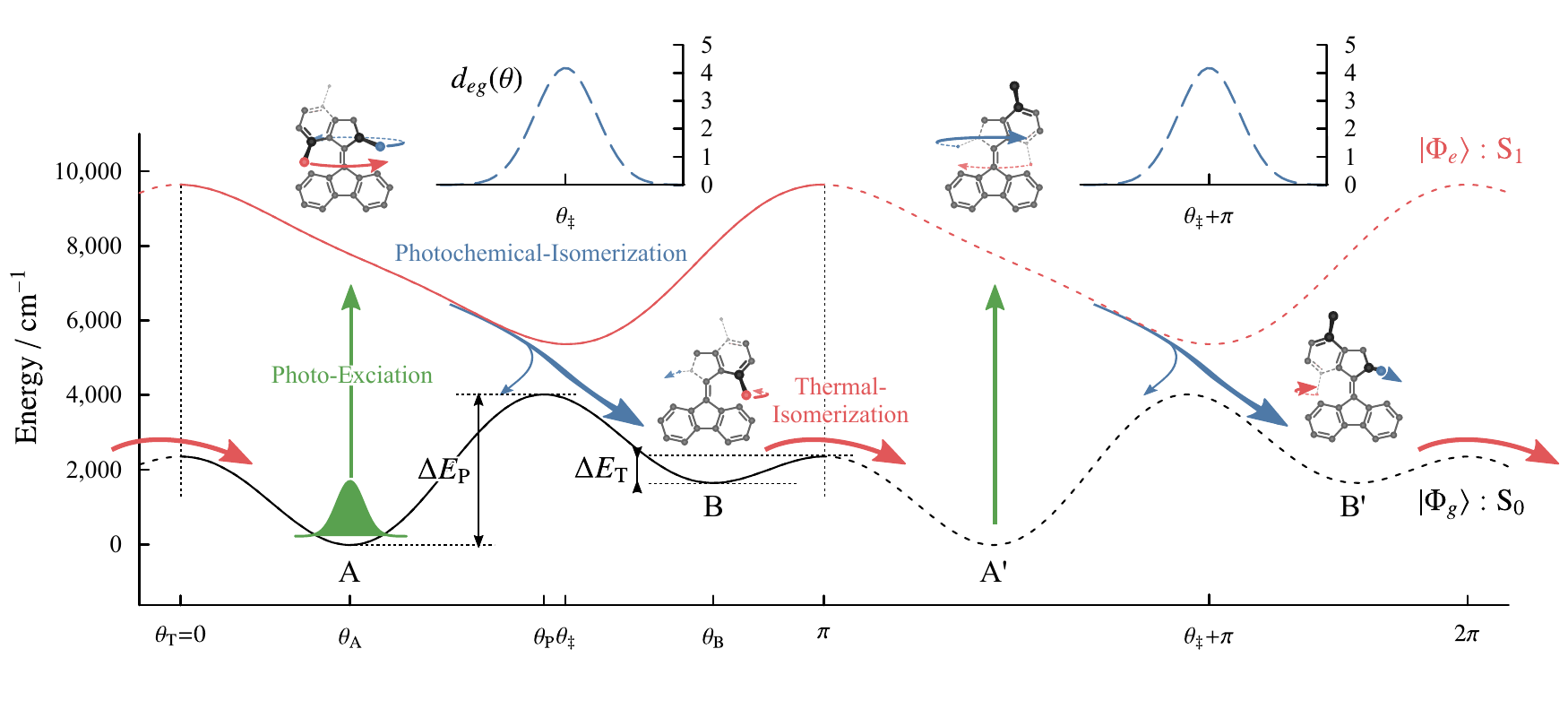}
  \caption{
    The adiabatic BO PESs and NAC (top insets) as a function of $\theta $.
    The black and red curves represent the ground ($\mathrm{S}_{0}$) and excited ($\mathrm{S}_{1}$) BO PESs, respectively.
    The thermal stable state and photo-product state are labeled by ``A'' and ``B''.
    Molecular structures of a typical two-step photo-driven molecular rotary motor (9-(2,4,7-trimethyl-2,3-dihydro-1H-inden-1ylidene)-9H-fluorene \cite{kazaryan2010jpca, kazaryan2011jctc, hu2016jcc}) are also depicted to illustrate the relationship between the present model and an actual molecular system.
  }
  \label{fig:PES} 
\end{figure*} 
We consider a molecular rotary motor system described by the two adiabatic electronic states, $|\Phi _{g}(\theta )\rangle $  and $|\Phi _{e}(\theta )\rangle $, which correspond to the $\mathrm{S}_{0}$ and $\mathrm{S}_{1}$ states.
Here, we have introduced the dimensionless variable $\theta $ ($0\leq \theta <2\pi $) to characterize a single rotation of the motor system.
Although this rotation process should be described by a set of dihedral angles representing the molecular configuration (e.g.~three dihedral angles $\alpha $, $\beta $ and $\theta $ as illustrated in Ref.~\onlinecite{kazaryan2010jpca}), here we assume that $\theta $ is treated as an ordinary coordinate/angle in order to describe the complex dynamics in a simple manner.
The system Hamiltonian is expressed as (see Appendix \ref{sec:hamiltonian-non-adiabatic})
\begin{align}
  \bm{H}(p_{\theta },\theta )&=\frac{\hbar }{2I_{\theta }}\Bigl(\hat{p}_{\theta }-i\bm{d}(\theta )\Bigr)^{2}+\bm{U}(\theta ),
  \label{eq:system-Hamiltonian-matrix} 
\end{align} 
where $p_{\theta }$ is the dimensionless conjugate momentum of $\theta $, and $I_{\theta }$ is the ``mass/moment of inertia'' of $\theta $, in which $I_{\theta }^{-1}$ has the dimension of frequency.
Here and hereafter, we employ the adiabatic matrix representation of the system Hamiltonian as $\{\bm{H}(p_{\theta },\theta )\}_{ab}\equiv \langle \Phi _{a}(\theta )|\hat{H}(p_{\theta },\theta )|\Phi _{b}(\theta )\rangle $ for $a,b=g,e$.
The matrix elements of the Born-Oppenheimer (BO) PESs are defined as $\{\bm{U}(\theta )\}_{ab}\equiv U_{a}(\theta )\delta _{ab}$, whereas those of the NACs are defined as $\{\bm{d}(\theta )\}_{a\neq b}\equiv d_{ab}(\theta )$ and $\{\bm{d}(\theta )\}_{aa}=0$.
%% new paragraph
While we can study four-step or multi-step molecular rotary motor systems \cite{feringa2011book}, here we consider a two-step molecular rotary motor for simplicity.
A typical two-step photo-driven molecular rotary motor is described by two successive processes that lead to unidirectional rotation:
(i) fast photoisomerization from $\mathrm{A}$ (the minimum state) to $\mathrm{B}$ (the meta-stable state) via photo-excitation, then (ii) slow thermalization from $\mathrm{B}$ to $\mathrm{A}'$ (another minimum state).

Although our approach can handle any form of the BO PESs and NACs, including those calculated from ab initio quantum chemical calculations, here we consider model PESs to capture characteristic features of the molecular motor system in a simple manner.
We thus consider the BO PESs in asymmetric periodic forms, expressed as
\begin{subequations}
  \begin{align}
    \begin{split}
      U_{g}(\theta )&=C_{1g}\sin 2(\theta +D_{1g})\\
      &\quad +C_{2g}\sin 4(\theta +D_{2g})+E_{g}
    \end{split}
    \intertext{and}
    \begin{split}
      U_{e}(\theta )&=C_{1e}\sin 2(\theta +D_{1e})\\
      &\quad +C_{2e}\sin 4(\theta +D_{2e})+E_{e}.
    \end{split}
  \end{align}
  \label{eq:pes}
\end{subequations}
Here, $C_{1/2,a}$, $D_{1/2,a}$, and $E_{a}$ for $a=g,e$ are the parameters for the two BO PESs.
The locations of the global minimum $\mathrm{A}$ and meta-stable states $\mathrm{B}$ are denoted by $\theta _{\mathrm{A}}$ and $\theta _{\mathrm{B}}$, respectively, while the locations of the global- and local-maximum of the ground PES that determine the transient states are denoted by $\theta _{\mathrm{P}}$ and $\theta _{\mathrm{T}}$, respectively.
Then the photoisomerization and thermalization are described as the movement of a wavepacket $\theta _{\mathrm{A}}\rightarrow \theta _{\mathrm{P}}\rightarrow \theta _{\mathrm{B}}$ and $\theta _{\mathrm{B}}\rightarrow \theta _{\mathrm{T}}\rightarrow \theta _{\mathrm{A}}+\pi $, respectively.
We introduce the vibrational frequency of the ground BO PES near each point, defined as
\begin{subequations}
  \begin{align}
    \hbar \Omega _{\mathrm{A}/\mathrm{B}}&\equiv \sqrt {+\frac{\hbar }{I_{\theta }}\left.\frac{\partial ^{2}U_{g}(\theta )}{\partial \theta ^{2}}\right|_{\theta =\theta _{\mathrm{A}/\mathrm{B}}}}
    \intertext{and}
    \hbar \Omega _{\mathrm{P}/\mathrm{T}}&\equiv \sqrt {\mathstrut -\frac{\hbar }{I_{\theta }}\left.\frac{\partial ^{2}U_{g}(\theta )}{\partial \theta ^{2}}\right|_{\theta =\theta _{\mathrm{P}/\mathrm{T}}}}.
  \end{align}
\end{subequations}
%% For convenience, we choose the value of $I_{\theta }$ equal to the vibrational motion of the hindered rotation around $\mathrm{A}$, and set the scale parameter $\Omega $ in Eq.~\eqref{eq:pes} in such a way that $\Omega _{\mathrm{A}}=\omega _{0}$.
The NACs between $|\Phi _{g}\rangle $ and $|\Phi _{e}\rangle $ are expressed using periodic Gaussian functions as
\begin{align}
  d_{eg}(\theta )&=-d_{ge}(\theta )
  =\mathcal{N}\sum _{m=-\infty }^{\infty }e^{-(\theta -\theta _{\ddagger }+m\pi )^{2}/2\sigma _{\ddagger }^{2}}.
\end{align}
Here $\theta _{\ddagger }$ and $\sigma _{\ddagger }$ are the center and width of the crossing region of the diabatic PESs, where the non-adiabatic transition occurs.
The normalization factor $\mathcal{N}$ is set to realize that
\begin{align}
  \int _{0}^{\pi }\!d\theta \,d_{eg}(\theta )=\frac{\pi }{2},
\end{align}
which means the two adiabatic bases are exchanged in the half period, $|\Phi _{e}(\theta +\pi )\rangle \propto |\Phi _{g}(\theta )\rangle $.

\begin{table}
  \centering
  \caption{ The parameter values for the present model. }
  \begin{tabular}{c@{~}r@{~}lc@{~}r@{~}l}
    \hline
    \hline
    Symbol & \multicolumn{2}{c}{Value} & Symbol & \multicolumn{2}{c}{Value}\\
    \hline
    \hline
    $C_{1g}$ & $-1,200$&$\mathrm{cm}^{-1}$ & $C_{1e}$ & $2,000$&$\mathrm{cm}^{-1}$\\
    $D_{1g}$ & $0.27$&       & $D_{1e}$ & $0.63$&\\
    $C_{2g}$ & $1,100$&$\mathrm{cm}^{-1}$ & $C_{2e}$ & $400$&$\mathrm{cm}^{-1}$\\
    $D_{2g}$ & $0.27$&       & $D_{2e}$ & $0.63$&\\
    $E_{g}$ & $2,000$&$\mathrm{cm}^{-1}$  & $E_{e}$ & $7,500$&$\mathrm{cm}^{-1}$\\
    \hline
    $\theta _{\ddagger }$ & $1.88$&$$ & $\sigma _{\ddagger }$ & $0.15$&\\
    $\hbar /I_{\theta }$ & $48.6$ &$\mathrm{cm}^{-1}$ & & &\\
    \hline
    \hline
  \end{tabular}
  \label{tab:parameters}
\end{table}
The parameter values of the PESs and NACs we employed are listed in Table~\ref{tab:parameters}.
The corresponding BO PESs and NACs are illustrated in Fig.~\ref{fig:PES}.
The barrier heights of the photoisomerization and thermalization are expressed as $\Delta E_{\mathrm{P}}=U_{g}(\theta _{\mathrm{P}})-U_{g}(\theta _{\mathrm{A}})$ and $\Delta E_{\mathrm{T}}=U_{g}(\theta _{\mathrm{T}})-U_{g}(\theta _{\mathrm{B}})$, respectively.
For these parameters, we have $\theta _{\mathrm{A}}=0.83$, $\theta _{\mathrm{B}}=2.60$, $\theta _{\mathrm{P}}=1.77$, $\theta _{\mathrm{T}}=0.00$, and $\Delta E_{\mathrm{P}}=4,034~\mathrm{cm}^{-1}$ and $\Delta E_{\mathrm{T}}=706~\mathrm{cm}^{-1}$.
The position of the minimum of the excited BO PES is $\theta _{e}=\theta _{\ddagger }=1.88$.
The value of $I_{\theta }$ is set to realize $\hbar \Omega _{\mathrm{A}}=100~\mathrm{cm}^{-1}$, and then the others are evaluated as $\hbar \Omega _{\mathrm{P}}=100~\mathrm{cm}^{-1}$ and $\hbar \Omega _{\mathrm{B}}=\hbar \Omega _{\mathrm{T}}=79.7~\mathrm{cm}^{-1}$.
Note that, we chose the shift parameters of the ground BO PES, $D_{1g}$ and $D_{2g}$, to set the local-maximum of the ground state potential at the boundary position, $\theta _{\mathrm{T}}\simeq 0$.
Our choice of these parameters exhibits a lower barrier height for thermalization, $\Delta E_{\mathrm{T}}$, than that of actual molecular motors systems.
This low barrier causes a faster thermalization than in actual systems and this allow us to suppress the computational costs.
However, we think these values are sufficient to capture the qualitative feature of ratchet dynamics of a photo-driven molecular motor system, where the timescales of the photoisomerization and thermalization processes are well separated.

\subsection{Multi-state low-temperature quantum Smoluchowski equations}
In order to study the thermal activation and deactivation processes near $\mathrm{B}$, we include environmental effects arising from the other molecular modes and the solvent modes.
We describe these as a heat bath, consisting of a set of harmonic oscillators.
The total Hamiltonian is then given by
\begin{align}
  \hat{H}_{\mathrm{tot}}(p_{\theta },\theta )&=\hat{H}(p_{\theta },\theta )+\hat{H}_{\mathrm{B}}(\vec{p},\vec{x};\theta ),
  \label{eq:total-Hamiltonian}
\end{align}
where
\begin{align}
  \hat{H}_{\mathrm{B}}(\vec{p},\vec{x};\theta )&=\sum _{\xi }\frac{\hbar \omega _{\xi }}{2}\left[\hat{p}_{\xi }^{2}+\left(x_{\xi }-\frac{g_{\xi }}{\omega _{\xi }}\theta \right)^{2}\right],
  \label{eq:bath}
\end{align}
and $\omega _{\xi }$, $x_{\xi }$, $p_{\xi }$, and $g_{\xi }$ represent the vibrational frequency, dimensionless coordinate, conjugate momentum, and system-bath coupling strength of the $\xi $th bath mode, respectively.
We denote the set of bath coordinates and momenta as $\vec{x}\equiv (\dots ,x_{\xi },\dots )$ and $\vec{p}\equiv (\dots ,p_{\xi },\dots )$, respectively.
The collective bath coordinate $\hat{X}\equiv \sum _{\xi }g_{\xi }\hat{x}_{\xi }$ acts as a source of noise for the motion along the PESs of the system.
This noise is characterized by the spectral density defined as
\begin{align}
  \mathcal{J}(\omega )\equiv \pi \sum _{\xi }\frac{g_{\xi }^{2}}{2}\delta (\omega -\omega _{\xi }).
  \label{eq:spectral-density}
\end{align}
In this paper, we employ the Ohmic spectral density,
\begin{align}
  \mathcal{J}(\omega )&=I_{\theta }\zeta \omega ,
\end{align}
which has been used frequently in quantum/classical models to study the dynamics under strong friction \cite{ankerhold2001prl, maier2010pre2, weiss2011book}.
Because the details of the phase-space dynamics of the wavepacket are not important for the present consideration, in particular for the thermalization process, we employ the strong friction case, $\zeta \gg \Omega _{\mathrm{A}/\mathrm{B}}$.

The dynamics of the above multi-state model with the bath, can be expressed by the reduced density matrix,
\begin{align}
  \{\bm{\rho }(\theta ,\theta ',t)\}_{ab}&=\rho _{ab}(\theta ,\theta ',t)\notag\\
  &\equiv \Bigl(\langle \theta |{\otimes }\langle \Phi _{a}(\theta )|\Bigr)\mathrm{Tr}_{\mathrm{B}}\{\hat{\rho }_{\mathrm{tot}}(t)\}\Bigl(|\Phi _{b}(\theta ')\rangle {\otimes }|\theta '\rangle \Bigr),
\end{align}
Here $\hat{\rho }_{\mathrm{tot}}(t)$ is the density operator of the whole system and $\mathrm{Tr}_{\mathrm{B}}\{\dots \}$ represents the trace operation for the bath space, $\vec{x}$.
The coordinate basis is expressed as $|\theta \rangle $.
In the strong friction case, the quantum coherence among the reduced coordinate space, $\bm{\rho }(\theta ,\theta ',t)$ for $\theta \neq \theta '$, vanishes \cite{grabert1988pr, philip2000adp, ikeda2018jctc}.
Then the state of the system is described by the probability distribution,
\begin{align}
  \{\bm{f}(\theta ,t)\}_{ab}=f_{ab}(\theta ,t)&\equiv \rho _{ab}(\theta ,\theta ,t).
\end{align}
Here, $f_{aa}(\theta ,t)$ is probability distribution of the $a$th adiabatic state, and $f_{ab}(\theta ,t)$ $(a\neq b)$ is the electronic coherence between the $a$th and $b$th adiabatic states.
It should be noted that, while the BO PESs and NACs are periodic, the system-bath interaction in Eq.~\eqref{eq:bath} has a non-periodic form.
When the quantum coherence of the periodic coordinate spreads over its period, the periodically invariant nature of the total wave function becomes important:
In such cases, we should employ a periodic system-bath model \cite{iwamoto2018jcp}.
In the overdamped case, however, such long-range quantum coherence vanishes and we can safely use the present model.

A single cycle of an actual two-step motor is achieved by repeating the photoisomerization/thermalization processes twice (i.e.~ $\mathrm{A}\rightarrow \mathrm{B}\rightarrow \mathrm{A}'$ and $\mathrm{A}'\rightarrow \mathrm{B}'\rightarrow \mathrm{A}$ in Fig.~\ref{fig:PES}).
Because the PESs and NACs in the first and second half cycles are identical, we regard $\mathrm{A}'$ as $\mathrm{A}$ and set the periodic boundary condition with the half cycle, $0\leq \theta <\pi $ to suppress computational costs.
Then we normalize $\bm{f}(\theta ,t)$ to be $1$ within this half cycle.

While we can ignore the quantum coherence among the $\theta $ space under strong friction, the quantum nature of the environment is still important, because the classical treatment of the fluctuations causes the violation of the positivity of the probability distribution, $\bm{f}(\theta ,t)$.
As explained in Sec.~\ref{sec:introduction}, to carry out physically consistent and accurate calculations, we employ the MS-LT-QSE---this is a variant of the HEOM in overdamped conditions---to take into account the low-temperature quantum thermal fluctuations arising from the environment.

Hereafter, we introduce the Liouville space operations,
\begin{subequations}
  \begin{align}
    \bm{O}(\theta )^{\rightarrow }\bm{f}(\theta )&\equiv \bm{O}(\theta )\bm{f}(\theta ),\\
    \bm{O}(\theta )^{\leftarrow }\bm{f}(\theta )&\equiv \bm{f}(\theta )\bm{O}(\theta ),\\
    \intertext{and}
    \bm{O}(\theta )^{\times /\circ }&\equiv \bm{O}(\theta )^{\rightarrow }\mp \bm{O}(\theta )^{\leftarrow }
  \end{align}
\end{subequations}
for arbitrary matrix $\bm{O}(\theta )$.
In the MS-LT-QSE theory, the dynamics of the system is described using a set of equations in a hierarchical structure, expressed as
\begin{widetext}
  \begin{align}
    \begin{split}
      \frac{\partial }{\partial t}\bm{f}_{\vec{n}}(\theta ,t)&=-\left[\mathcal{E}+\sum _{k}^{K}n_{k}\gamma _{k}+\frac{1}{I_{\theta }\zeta }\Bigl(\mathcal{F}+\Xi _{K}\Bigr)\right]\bm{f}_{\vec{n}}(\theta ,t)
      -\sum _{k}^{K}\Phi \bm{f}_{\vec{n}+\vec{e}_{k}}(\theta ,t)-\frac{1}{I_{\theta }\zeta }\sum _{k}^{K}n_{k}\gamma _{k}\Theta _{k}\bm{f}_{\vec{n}-\vec{e}_{k}}(\theta ,t),
    \end{split}
    \label{eq:ms-lt-qse}
  \end{align}
\end{widetext}
where $\vec{n}\equiv (\dots ,n_{k},\dots )$ is a multi-index whose components are all non-negative integers and $\vec{e}_{k}\equiv (0,\dots ,1,0,\dots )$ is the $k$th unit vector.
\begin{subequations}
  \begin{align}
    \mathcal{E}&\equiv \frac{i}{\hbar }\bm{U}(\theta )^{\times }
  \end{align}
  is the Liouvillian of the quantum von-Neumann equation of the electronic subspace at $q$, and
  \begin{align}
    \mathcal{F}&\equiv \frac{1}{2}\bm{F}(\theta )^{\circ }\left(\frac{\partial }{\partial \theta }+\bm{d}(\theta )^{\times }\right)+\frac{1}{2}\bm{A}(\theta )^{\circ }
    \label{eq:drift}
  \end{align}
  is the drift term which also takes into account the electronic transition processes.
\end{subequations}
\begin{subequations}
  Here, we have introduced the force acting on the adiabatic states as
  \begin{align}
    \begin{split}
      \bm{F}(\theta )&\equiv -\frac{1}{\hbar }\frac{\partial \bm{U}(\theta )}{\partial \theta }-\frac{1}{\hbar }\bigl[\bm{d}(\theta )\bm{U}(\theta )-\bm{U}(\theta )\bm{d}(\theta )\bigr]
      \label{eq:F}
    \end{split}
  \end{align}
  and its derivatives,
  \begin{align}
    \begin{split}
      \bm{A}(\theta )&\equiv +\frac{\partial \bm{F}(\theta )}{\partial \theta }+\bigl[\bm{d}(\theta )\bm{F}(\theta )-\bm{F}(\theta )\bm{d}(\theta )\bigr].
      \label{eq:A}
    \end{split}
  \end{align}
\end{subequations}
The relaxation operators in Eq.~\eqref{eq:ms-lt-qse} are defined by
\begin{subequations}
  \begin{align}
    \Phi &\equiv -\left(\frac{\partial }{\partial \theta }+\bm{d}(\theta )^{\times }\right)
    \label{eq:Phi},\\
    \Theta _{k}&\equiv -\frac{2\eta _{k}}{\beta \hbar }\Phi ,
    \intertext{and}
    \Xi _{K}&\equiv -\frac{1+\sum _{k}^{K}2\eta _{k}}{\beta \hbar }\Phi ^{2},
  \end{align}
\end{subequations}
where $\beta \equiv 1/k_{\mathrm{B}}T$ is the inverse temperature divided by the Boltzmann constant.

In this formalism, the coefficients $\nu _{k}$ and $\eta _{k}$ are introduced in Eq.~\eqref{eq:ms-lt-qse} to avoid the positivity problem and to have accurate quantum dynamics, and are chosen so as to realize the relation
\begin{align}
  n(\omega )+\frac{1}{2}\approx \frac{1}{\beta \hbar }\frac{1}{\omega }+\sum ^{K}_{k}\frac{2\eta _{k}}{\beta \hbar }\frac{\omega }{\omega ^{2}+\nu _{k}^{2}}
  \label{eq:temperature-kernel}
\end{align}
for the finite $K\geq 0$.
Here, $n(\omega )\equiv (e^{\beta \hbar \omega }-1)^{-1}$ is the BE distribution function.
The first term on the right-hand side in Eq.~\eqref{eq:temperature-kernel} corresponds to the classical term, while the remaining terms represent the QLT correction.
If we consider the case $K\rightarrow \infty $, $\nu _{k}$ should be the $k$th Matsubara frequency, $\tilde{\nu }_{k}\equiv 2\pi k/\beta \hbar $, and then $\eta _{k}=1$.
The quantum fluctuations are essentially non-Markovian, and this is an important consequence of the quantum fluctuation-dissipation (QFD) theorem \cite{tanimura2006jpsj, weiss2011book}.
The MS-LT-QSE theory treats the effect of the fluctuations in a manner consistent with the QFD theorem.
The multi-index $\vec{n}$ represents the index of the hierarchy, and the first hierarchical element, $\bm{f}_{\vec{0}}(\theta ,t)$, has a physical meaning as the probability distribution (i.e.~$\bm{f}_{\vec{0}}(\theta ,t)=\bm{f}(\theta ,t)$).
The rest of the hierarchical elements, $\bm{f}_{\vec{n}\neq \vec{0}}(\theta ,t)$, allow the treatment of non-Markovian QLT effects.

In order to carry out numerical calculations, we set $K$ to a finite value.
This makes the expansion in Eq.~\eqref{eq:temperature-kernel} adjustable, while the integer $K$ determines the accuracy of the calculations.
Here, we employ the Pad\'{e} spectral decomposition (PSD) [N-1/N] scheme to enhance the computational efficiency of calculations while maintaining the numerical accuracy.
Then we can truncate $f_{\vec{n}}(\theta ,t)$ for $\vec{n}\in \mathbb{N}^{K}$ into finite elements to carry out numerical calculations with sufficient accuracy.
In this paper, we employ hierarchical elements $\bm{f}_{\vec{n}}(\theta ,t)$ which satisfy the relation $\Delta _{\vec{n}}\Omega _{\mathrm{A}}/\gamma _{\vec{n}}>\delta _{\mathrm{tol}}$, where $\delta _{\mathrm{tol}}$ is the tolerance of the truncation and $\gamma _{\vec{n}}\equiv \sum _{k}^{K}n_{k}\nu _{k}$ and
\begin{align}
  \Delta _{\vec{n}}&\equiv \prod _{k}^{K}\frac{1}{n_{k}!}\left(\frac{\eta _{k}}{\eta _{K}}\right)^{n_{k}}.
\end{align}
The condition $\delta _{\mathrm{tol}}=0$ implies that we employ all elements.
The validity of the truncation is justified by adjusting the value of $\delta _{\mathrm{tol}}$.

In the high-temperature or classical limit (i.e.~$K=0$ in Eq.~\eqref{eq:temperature-kernel}), the set of Eq.~\eqref{eq:ms-lt-qse} become a Markovian equation,
\begin{align}
  \frac{\partial }{\partial t}\bm{f}(\theta ,t)&=-\left[\mathcal{E}+\frac{\omega _{0}}{\zeta }\left(\mathcal{F}+\Xi _{0}\right)\right]\bm{f}(\theta ,t).
  \label{eq:ms-se}
\end{align}
This is the multi-state Smoluchowski equation (MS-SE) \cite{ikeda2018jctc}.
When the diabatic PESs of the system are harmonic, the MS-SE reduces to the Zusman equation (ZE) that is frequently used for investigations of electron transport problems \cite{zusman1980cp, garg1985jcp, jung2002jcp, shi2009jcp2, shi2009jcp}.
Note that, in the original form of the ZE in Ref.~\onlinecite{zusman1980cp}, the forces described in Eq.~\eqref{eq:F} is approximated as state-independent.
State-dependent force term is introduced in Ref.~\onlinecite{garg1985jcp}.

For a system with a single electronic state, all matrices in the electronic subspace reduce to scalar functions (e.g.~$\bm{f}(\theta ,t)\rightarrow f(\theta ,t)$) that leads Eq.~\eqref{eq:ms-se} to be the classical Smoluchowski equation,
\begin{align}
  \frac{\partial }{\partial t}f(\theta ,t)&=\frac{1}{I_{\theta }\zeta }\frac{\partial }{\partial \theta }\left(-F(\theta )+\frac{1}{\beta \hbar }\frac{\partial }{\partial \theta }\right)f(\theta ,t).
  \label{eq:se}
\end{align}
Because we have employed the dimensionless coordinate $\theta $, the Planck constant appears in this classical equation.

\subsection{Photoexcitation}
\label{sec:photoexcitation}
Wavepacket dynamics with non-adiabatic transitions and thermalization processes are described using Eq.~\eqref{eq:ms-lt-qse}.
The photo-excitation process caused by an electric field is described by the replacement \cite{mukamel1999book},
\begin{align}
  \bm{U}(\theta )\rightarrow \bm{U}(\theta )-E(t)\bm{\mu }(\theta )
  \label{eq:PES-replacement},
\end{align}
where $E(t)$ is the electric field as a function of time $t$ and $\left\{\bm{\mu }(\theta )\right\}_{ab}=\langle \Phi _{e}(\theta )|\hat{\mu }|\Phi _{b}(\theta )\rangle $ is the dipole element.
Then Green's function of the total system is expressed as
\begin{align}
  \mathcal{G}_{\mathrm{tot}+\mathrm{p}}(t,t_{0})\equiv \mathop{\exp }_{\leftarrow }\left[-\int _{t_{0}}^{t}\!ds\,\Bigl(\mathcal{L}_{\mathrm{tot}}+\mathcal{L}_{\mathrm{ p}}(s)\Bigr)\right],
  \label{eq:green-function-tot-p}
\end{align}
where
\begin{subequations}
  \begin{align}
    \mathcal{L}_{\mathrm{tot}}&\equiv \frac{i}{\hbar }\hat{H}_{\mathrm{tot}}^{\times }
    \intertext{and}
    \mathcal{L}_{\mathrm{p}}(t)&\equiv -E(t)\frac{i}{\hbar }\hat{\mu }^{\times }
  \end{align}
\end{subequations}
are the Liouvillian operator of the total system and pulse interaction, respectively, and $\exp _{\leftarrow }$ is the time ordered exponential.

Experimentally, several schemes have been developed to drive motor systems, e.g.~continuous or intermittent irradiation of laser \cite{feringa2011book}.
To simplify the analysis, here we consider the case that the motor system is driven by repeated pulses, and that each laser pulse is impulsive assuming that photoexcitation is much faster than the photoisomerization process.
Then we have
\begin{align}
  E(t)&=\sum _{n=-\infty }^{\infty }\bar{E}\Delta \tau \delta (t-n\tau _{\mathrm{p}}),
\end{align}
where $\tau _{\mathrm{p}}$ is the repetition time interval of the laser pulses and $\bar{E}$ and $\Delta t$ are the effective intensity and duration of each pulse.
Then Eq.~\eqref{eq:green-function-tot-p} is separated into the total system part and pulse interaction part, and we have
\begin{align}
  \mathcal{G}_{\mathrm{tot}+\mathrm{p}}(t,t_{0})\equiv \mathcal{G}_{\mathrm{tot}}(\Delta t)\mathcal{G}_{\mathrm{p}}\Bigl(\mathcal{G}_{\mathrm{tot}}(\tau _{\mathrm{p}})\mathcal{G}_{\mathrm{p}}\Bigr)^{N-1}\mathcal{G}_{\mathrm{ tot}}(\Delta t_{0}),
  \label{eq:green-function-tot-p-repetition}
\end{align}
where
\begin{subequations}
  \begin{align}
    \mathcal{G}_{\mathrm{tot}}(t-t_{0})&\equiv \exp \left[-\frac{i}{\hbar }\mathcal{L}_{\mathrm{tot}}(t-t_{0})\right]
    \label{eq:green-function-tot}
    \intertext{and}
    \mathcal{G}_{\mathrm{p}}&\equiv \exp \left(\frac{i}{\hbar }\bar{E}\Delta \tau \hat{\mu }^{\times }\right)
    \label{eq:green-function-p}
  \end{align}
\end{subequations}
are Green's function of the total system without pulse interactions and that for the impulsive pulse interaction, respectively.
The time interval is expressed as $t-t_{0}=\Delta t+(N-1)\tau _{\mathrm{p}}+\Delta t_{0}$, where $N$ is the number of pulses, $\Delta t_{0}$ and $\Delta t$ are the durations from $t_{0}$ to the first pulse interaction and from the last pulse interaction to $t$.

In the HEOM formalism, the total density operator is replaced by a set of hierarchical elements, $\{\bm{f}_{\vec{n}}(t)\}$, that describe the non-perturbative interactions between the system and bath \cite{tanimura2014jcp, tanimura2015jcp}.
Then Eq.~\eqref{eq:green-function-tot} is evaluated by integrating Eq.~\eqref{eq:ms-lt-qse}.
The pulse interaction, $(i/\hbar )\hat{\mu }^{\times }$ in Eq.~\eqref{eq:green-function-p}, is also evaluated from the potential terms in Eq.~\eqref{eq:ms-lt-qse} with the replacement $\bm{U}(\theta )\rightarrow -\bm{\mu }(\theta )$ (see~Eq.~\eqref{eq:PES-replacement}).
Therefore,
\begin{align}
  \begin{split}
    (i/\hbar )\hat{\mu }^{\times }\rho _{\mathrm{tot}}(t)&\rightarrow \mathcal{E}_{\mu }\bm{f}_{\vec{n}}(\theta ,t)+\frac{\omega _{0}}{\zeta }\mathcal{F}_{\mu }\bm{f}_{\vec{n}}(\theta ,t),
  \end{split}
  \label{eq:ms-lt-qse-p}
\end{align}
where
\begin{subequations}
  \begin{align}
    \mathcal{E}_{\mu }&\equiv \frac{i}{\hbar }\bm{\mu }(\theta )^{\times }
  \end{align}
  represents the vertical transition among the adiabatic states, and
  \begin{align}
    \mathcal{F}_{\mu }&\equiv \frac{1}{2}\bm{F}_{\mu }(\theta )^{\circ }\left(\frac{\partial }{\partial \theta }+\bm{d}(\theta )^{\times }\right)+\frac{1}{2}\bm{A}_{\mu }(\theta )^{\circ }
  \end{align}
  is the force term driven by the pulse interaction.
\end{subequations}
Here, $\bm{F}_{\mu }(\theta )$ and $\bm{A}_{\mu }(\theta )$ are defined by the replacement $\bm{U}(\theta )\rightarrow \bm{\mu }(\theta )$ with Eqs.~\eqref{eq:F} and \eqref{eq:A}, respectively.

For simplicity, we assume that the dipole operator has only off-diagonal components, i.e.~$\hat{\mu }=\mu \left(|\Phi _{e}\rangle \langle \Phi _{g}|+|\Phi _{g}\rangle \langle \Phi _{e}|\right)$, where $\mu $ is the dipole strength that is independent from $q$ in the adiabatic representation.
Note that, although $\bm{F}_{\mu }(\theta )$ is non-zero and $\theta $-dependent under this condition due to the second term in Eq.~\eqref{eq:F}, we omit the second term in Eq.~\eqref{eq:ms-lt-qse-p} because the effect of non-vertical transitions is minor.
Then the action of the $\mathcal{G}_{\mathrm{p}}$ to $\{\bm{f}_{\vec{n}}(t)\}$ can be rewritten as (see Appendix \ref{sec:pulse-green-function})
\begin{widetext}
  \begin{align}
    \mathcal{G}_{\mathrm{p}}\bm{f}_{\vec{n}}(\theta ,t)
    &\simeq (1-\alpha _{\mathrm{E}})
    \begin{pmatrix}
      f_{gg} & f_{ge}\\
      f_{eg} & f_{ee}\\
    \end{pmatrix}_{\vec{n}}
    +\alpha _{\mathrm{E}}
    \begin{pmatrix}
      f_{ee} & f_{eg}\\
      f_{ge} & f_{gg}\\
    \end{pmatrix}_{\vec{n}}
    +i\sin \varphi _{\mathrm{p}}\cos \varphi _{\mathrm{p}}
    \begin{pmatrix}
      f_{eg}-f_{ge} & f_{ee}-f_{gg}\\
      f_{gg}-f_{ee} & f_{ge}-f_{eg}\\
    \end{pmatrix}_{\vec{n}},
    \label{eq:photoexcitation}
  \end{align}
\end{widetext}
where $\varphi _{\mathrm{p}}\equiv \bar{E}\mu \Delta \tau /\hbar $ and $\alpha _{\mathrm{E}}\equiv \sin ^{2}\varphi _{\mathrm{p}}$ ($0\leq \alpha _{\mathrm{E}}\leq 1$), and we simplified the notation with $f_{ab}=f_{ab}(\theta ,t)$ for $a,b=g,e$.
As the above equation indicates, the constant $\alpha _{\mathrm{E}}$ represents the excitation ratio of the molecules by a single pulse interaction, and the excitation pulses exchange the populations in the $|\Phi _{g}\rangle $ and $|\Phi _{e}\rangle $ states completely for $\alpha _{\mathrm{E}}=1$.

\subsection{Linear/non-linear optical spectroscopies}
The optical response function for the total system, $\hat{\rho }_{\mathrm{tot}}(0)$, is written as \cite{tanimura2006jpsj}
\begin{align}
  \begin{split}
    R(\tau )&\equiv \mathrm{Tr}\biggl\{\hat{\mu }\mathcal{G}_{\mathrm{tot}}(\tau )\frac{i}{\hbar }\hat{\mu }^{\times }\hat{\rho }_{\mathrm{tot}}(0)\biggr\}.
  \end{split}
  \label{eq:response}
\end{align}
In the case that $\hat{\rho }_{\mathrm{tot}}(0)=\hat{\rho }_{\mathrm{tot}}^{\mathrm{eq}}$ is the thermal equilibrium state, Eq.~\eqref{eq:response} corresponds to the linear response, $R_{\mathrm{A}}(\tau )$.
When $\hat{\rho }_{\mathrm{tot}}(0)=\mathcal{G}_{\mathrm{tot}}(t)\hat{\rho }_{\mathrm{tot}}(t_{0})$ is time-dependent, Eq.~\eqref{eq:response} corresponds to a transient response, $R_{\mathrm{TA}}(\tau ,t)$.
In the case that the initial state is perturbed by laser pulses as $\hat{\rho }_{\mathrm{tot}}(0)=\mathcal{G}_{\mathrm{tot}}(t)[(i/\hbar )\hat{\mu }^{\times }]^{2}\hat{\rho }_{\mathrm{tot}}^{\mathrm{eq}}$, Eq.~\eqref{eq:response} corresponds to the pump-probe response, $R_{\mathrm{PP}}(\tau ,t)$.
Because Eq.~\eqref{eq:response} is the first-order response function of the laser excitation from $\hat{\rho }_{\mathrm{tot}}(0)$, we can calculate the absorption spectrum using the Fourier transform as
\begin{align}
  I_{\mathrm{A}/\mathrm{TA}/\mathrm{PP}}(\omega ,t)&\equiv \omega \mathrm{Im}\int _{0}^{\infty }\!d\tau \,e^{i\omega \tau }R_{\mathrm{A}/\mathrm{TA}/\mathrm{PP}}(\tau ,t).
  \label{eq:response-spectrum}
\end{align}
We further evaluate the third-order rephasing/non-rephasing responses using the rotating wave approximation for the pulse interactions as \cite{ishizaki2006jcp, ishizaki2007jpca}
\begin{widetext}
  \begin{subequations}
    \begin{align}
      R_{\mathrm{R}}(t_{3},t_{2},t_{1})&\equiv \mathrm{Tr}\biggl\{\hat{\mu }\mathcal{G}_{\mathrm{tot}}(t_{3})\frac{i}{\hbar }\hat{\mu }_{\rightarrow }^{\times }\mathcal{G}_{\mathrm{tot}}(t_{2})\frac{i}{\hbar }\hat{\mu }_{\rightarrow }^{\times }\mathcal{G}_{\mathrm{tot}}(t_{1})\frac{i}{\hbar }\hat{\mu }_{\leftarrow }^{\times }\hat{\rho }_{\mathrm{tot}}(0)\biggr\}
      \label{eq:r-response}
      \intertext{and}
      R_{\mathrm{NR}}(t_{3},t_{2},t_{1})&\equiv \mathrm{Tr}\biggl\{\hat{\mu }\mathcal{G}_{\mathrm{tot}}(t_{3})\frac{i}{\hbar }\hat{\mu }_{\rightarrow }^{\times }\mathcal{G}_{\mathrm{tot}}(t_{2})\frac{i}{\hbar }\hat{\mu }_{\leftarrow }^{\times }\mathcal{G}_{\mathrm{tot}}(t_{1})\frac{i}{\hbar }\hat{\mu }_{\rightarrow }^{\times }\hat{\rho }_{\mathrm{tot}}(0)\biggr\},
      \label{eq:nr-response}
    \end{align}
  \end{subequations}
\end{widetext}
where $\hat{\mu }_{\rightarrow }\equiv \mu |\Phi _{e}\rangle \langle \Phi _{g}|$ and $\hat{\mu }_{\leftarrow }\equiv \mu |\Phi _{g}\rangle \langle \Phi _{e}|$.
Then, the two-dimensional (2D) correlation spectrum is expressed as
\begin{align}
  I_{\mathrm{2D}}(\omega _{3},t_{2},\omega _{1})&\equiv I_{\mathrm{R}}(\omega _{3},t_{2},-\omega _{1})+I_{\mathrm{NR}}(\omega _{3},t_{2},\omega _{1}),
\end{align}
where
\begin{align}
  \begin{split}
    I_{\mathrm{R}/\mathrm{NR}}(\omega _{3},t_{2},\omega _{1})&\equiv \omega _{3}\mathrm{Im}\int _{0}^{\infty }\!dt_{3}\,e^{i\omega _{3}t_{3}}\int _{0}^{\infty }\!dt_{1}\,e^{i\omega _{1}t_{1}}\\
    &\quad \times R_{\mathrm{R}/\mathrm{NR}}(t_{3},t_{2},t_{1}).
  \end{split}
  \label{eq:r-nr-spectrum}
\end{align}
Here, we have introduced the prefactor $\omega _{3}$ in Eq.~\eqref{eq:r-nr-spectrum} to compare with Eq.~\eqref{eq:response-spectrum}.

In the HEOM formalism, Green's function in Eqs.~\eqref{eq:response}, \eqref{eq:r-response}, and \eqref{eq:nr-response} are evaluated by integrating Eq.~\eqref{eq:ms-lt-qse}.
The trace calculation of the total system is carried out as the trace of the zeroth hierarchical element, $\bm{f}_{\vec{0}}(t)$.

\subsection{Flux of rotary motor system}
We introduce the probability distribution in $q$-space expressed as
\begin{align}
  f(\theta ,t)&\equiv \sum _{a}f_{aa}(\theta ,t).
\end{align}
By inserting this into Eq.~\eqref{eq:ms-lt-qse} and by comparing with the continuity equation of the probability distribution,
\begin{align}
  \frac{\partial }{\partial t}f(\theta ,t)=-\frac{\partial }{\partial \theta }j(\theta ,t),
\end{align}
we obtain the flux $j(\theta ,t)$ expressed as
\begin{widetext}
  \begin{align}
    \begin{split}
      j(\theta ,t)&=\frac{1}{I_{\theta }\zeta }\sum _{ab}F_{ab}(\theta )f_{\vec{0},ba}(\theta ,t)
      -\frac{1}{I_{\theta }\zeta }\frac{1+\sum _{k}^{K}2\eta _{k}}{\beta \hbar }\sum _{a}\frac{\partial }{\partial \theta }f_{\vec{0},aa}(\theta )
      -\sum _{k}^{K}\sum _{a}f_{\vec{e}_{k},aa}(\theta ,t).
    \end{split}
    \label{eq:flux}
  \end{align}
\end{widetext}
The first term represents the flux arising from the potential, while the second and third terms represent the contribution from the thermal fluctuations.
The net probability distribution flowing through the point $q$ during the time $t_{a}\leq t'\leq t_{b}$ is then expressed as
\begin{align}
  J(\theta ,t_{b},t_{a})&\equiv \int _{t_{a}}^{t_{b}}\!dt'\,j(\theta ,t'),
\end{align}
and the averaged flux is evaluated as $\bar{\jmath }(\theta ,t_{b},t_{a})\equiv J(\theta ,t_{b},t_{a})/(t_{b}-t_{a})$.
Using $J(\theta ,t_{b},t_{a})$ and $\bar{\jmath }(\theta ,t_{b},t_{a})$, we can quantify the average rotational speed of the molecular motor from numerical calculations.

\section{NUMERICAL RESULTS}
\label{sec:results}
We now present the numerical results.
In the following, all numerical calculations carried out to integrate Eq.~\eqref{eq:ms-lt-qse} were performed using the fourth-order low-storage Runge-Kutta method with a time step $\delta t=0.4\times 10^{-3}~\mathrm{ps}$ \cite{yan2017cjcp}.
The finite difference calculations for the $q$-derivatives in Eq.~\eqref{eq:ms-lt-qse} were performed using the first and second-order difference method with eighth-order accuracy \cite{fornberg1988mc}.
The time integrations of $j(\theta ,t)$ were performed using the trapezoidal rule, while the numerical integrations of $\bm{f}(\theta ,t)$ in the $q$-space were performed using the rectangle rule.
The mesh size in $q$-space was set to $N_{q}=64$ with ranges $0\leq q<\pi $.
The bath coupling parameter was set to $\zeta =2.5\,\omega _{0}$ (overdamped) and $T=300~\mathrm{K}$ ( $\beta \hbar \omega _{0}=0.48$ ).
The parameters of the QLT correction were set to $K=2$ with the tolerance $\delta _{\mathrm{tol}}=5\times 10^{-4}$ as explained below.
With these parameters, the PSD[N-1/N] coefficients are calculated as $\eta _{1}=1.03$, $\nu _{1}=6.31/\beta \hbar =1,316~\mathrm{cm}^{-1}$, $\eta _{2}=5.97$, and $\nu _{2}=19.5/\beta \hbar =4,066~\mathrm{cm}^{-1}$, and the number of total hierarchical elements is eight.
Under these conditions, the computational memory used for $\bm{f}_{\vec{n}}(\theta ,t)$ is $32~\mathrm{kB}$ in the double-precision floating-point number format.
The time integrations of the MS-LT-QSE were carried out by running a C++ program with the Intel Math Kernel Library (MKL) sparse Basic Linear Algebra Subprograms (BLAS) library.
For $1,000~\mathrm{ps}$ calculation using a single thread code running on the Intel Core i7-6650U central processing unit (CPU) in a note book computer, the computational time was approximately $20$ minutes.
For the same calculation using a parallelized code with four threads running on the Intel Xeon W-2125 CPU, the computational times was approximately $5$ minutes.
This demonstrated that a simulation for microsecond timescales---which corresponds to the thermalization of actual molecular motor systems---is possible in approximately two weeks with a notebook computer, although a computational time strongly depends on parameters of calculations, e.g.~electronic resonant frequencies of the system, the barrier heights of the PESs, temperature and system-bath coupling strength.
%% If necessary, we may also adapt our code to the Graphical Processer Unit (GPU).
%% \comment{The GPU code have already been developt.}

\subsection{Role of the QLT correction terms}
\label{sec:exam-qltc}
\begin{figure} \centering
\includegraphics[scale=\SingleColFigScale]{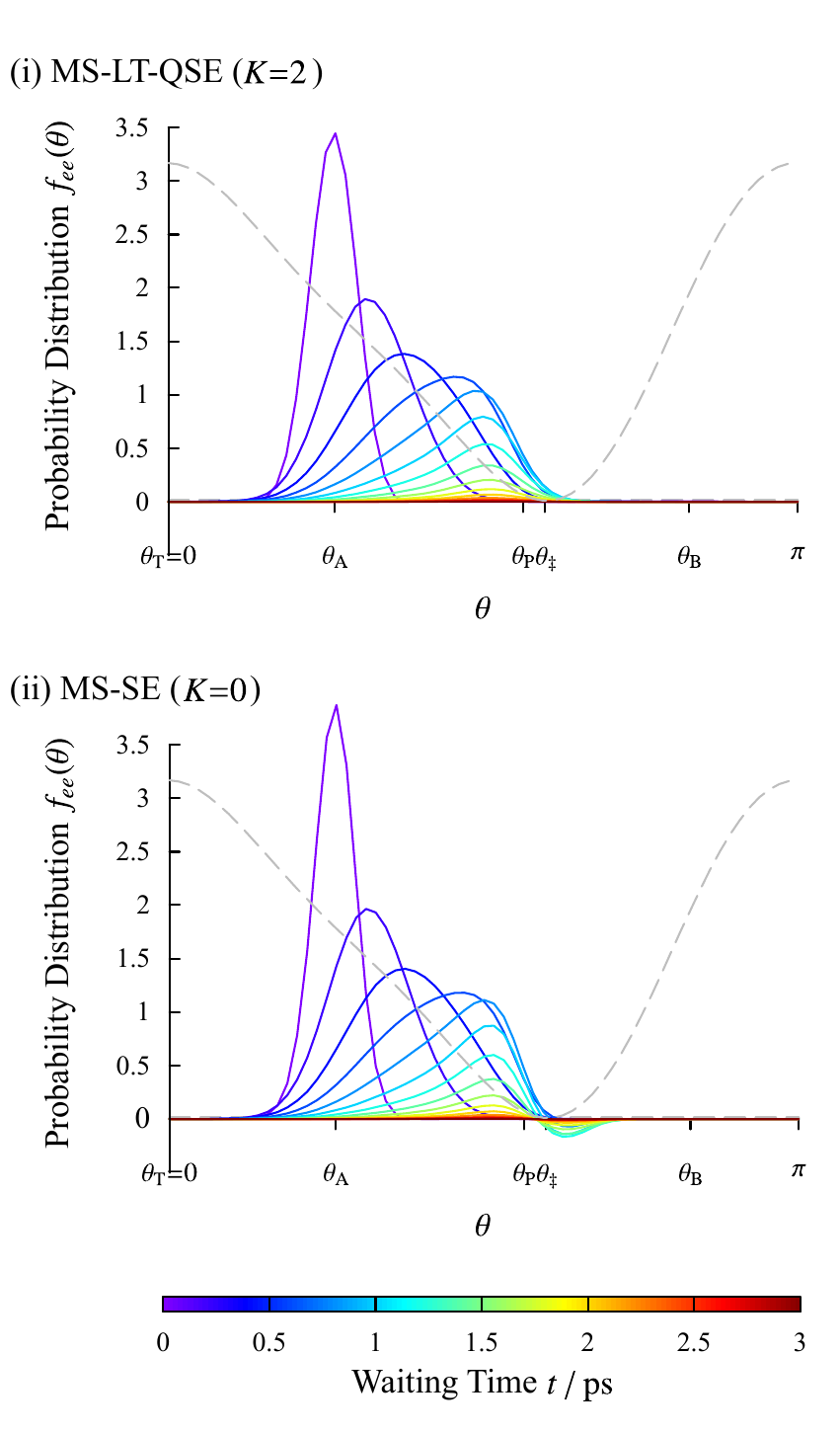}
\caption{
  Snapshots of wavepackets, $f_{ee}(\theta ,t)$, in the adiabatic excited state after the photoexcitation at $t=0$.
  The colors of the curves represent the different waiting time after the photoexcitation.
  The gray dotted curves represent the excited PES, $U_{e}(\theta )$, with an arbitrary unit.
  (i) The calculated results with the QLT correction ($K=2$) and (ii) the calculated results without the QLT correction ($K=0$; MS-SE).  }
\label{fig:wavepacket}
\end{figure}
To demonstrate the importance of the quantum low-temperature correction terms in the MS-LT-QSE theory, we first illustrate the snapshots of wavepackets created by the photoexcitation at $t=0$.
For this purpose, we prepared the initial state as the thermal equilibrium state of the system, which is obtained as the steady state solution of Eq.~\eqref{eq:ms-lt-qse}.
The excited wavepacket was then created by the laser pulse, described by Eq.~\eqref{eq:photoexcitation} with $\alpha _{\mathrm{E}}=1$ at $t=0$.
The time evolution of the photoinduced wavepackets was then computed by running the program for the MS-LT-QSE for $t>0$.

In Fig.~\eqref{fig:wavepacket}, the wavepacket calculations with and without the QLT correction are depicted.
The number of QLT correction terms is set to $K=2$ with the tolerance $\delta _{\mathrm{tol}}=5\times 10^{-4}$.
While the MS-LT-QSE theory predicts accurate wavepacket dynamics, the MS-SE theory exhibits a negative probability distribution near the crossing region $\sim \theta _{\ddagger }$ in the non-adiabatic transition period ($0.4~\mathrm{ps}\lesssim t\lesssim 3~\mathrm{ps}$), which is by no means physical.
This negative population in the non-adiabatic transition is due to the unrealistic Markovian assumption at low temperature, and is known as the positivity problem \cite{tanimura2006jpsj,ikeda2018jctc}:
The MS-SE theory is valid only in a high temperature regime, where the excited states are thermally well populated.
On the contrary, the MS-LT-QSE theory can predict dynamics accurately even at low temperature using a Smoluchowski-like equation, while the computational time is only $8$ times slower than the MS-SE case under the current parameters.
In the following, we set $K=2$ and $\delta _{\mathrm{tol}}=5\times 10^{-4}$ for all of the calculations.

Here, we note the following two points:
First, under the limit $K\rightarrow \infty $, the probability distribution, $\bm{f}(\theta ,t)$, may diverge in the present MS-LT-QSE theory \cite{ikeda2018jctc}.
Therefore, we need to fix the number of QLT correction terms, $K$, to a finite value.
In the present calculation, $K$ was chosen to be the minimal value which avoids the negative population.
Second, apart from the negative population, the results of the MS-LT-QSE ($K=2$) and MS-SE ($K=0$) are qualitatively similar and we may adopt the MS-SE in the present case.
The validity of MS-SE is dependent on the choice of parameters, however:
If the temperature is lower than the present calculation, or the energy gap of the PESs is higher than the present calculation, the MS-SE may cause serious error arising from the negative population.

\subsection{Photoisomerization}
\label{sec:photoisomerization}
We study the dynamical behavior of the non-adiabatic transition right after the photoexcitation.
The calculation of excited dynamics is same as the case of $K=2$ in Sec.~\ref{sec:exam-qltc}.
\begin{figure}
  \centering
  \includegraphics[scale=\SingleColFigScale]{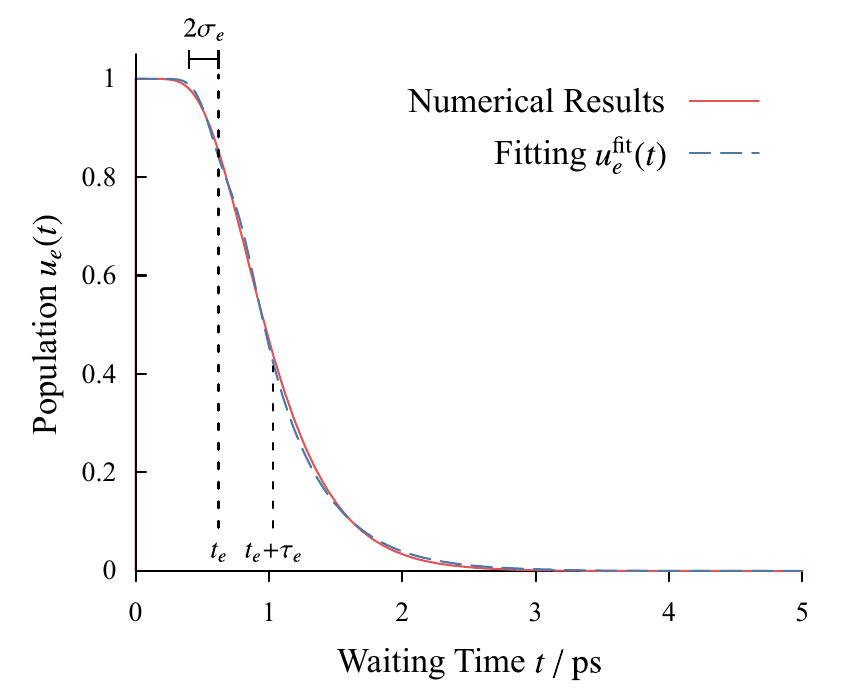}
  \caption{ The excited population $u_{e}(t)$ is plotted as a function of the time.
    The red solid curve represents the numerical result, whereas the blue dashed curve is the fitted result using Eq.~\eqref{eq:fitting-function} with the fitting parameters $t_{e}=0.62~\mathrm{ps}$, $\sigma _{e}=0.11~\mathrm{ps}$, and $\tau _{e}=0.41~\mathrm{ps}$.
  }
  \label{fig:nonadiabatic_transition}
\end{figure}
To extract the timescale of the non-adiabatic transition, we define the excited state population as
\begin{align}
  u_{e}(t)&\equiv \int _{0}^{\pi }\!d\theta \,f_{ee}(\theta ,t).
\end{align}
In Fig.~\ref{fig:nonadiabatic_transition}, we depict the calculated result and the fitted result using the function
\begin{align}
  u_{e}^{\mathrm{fit}}(t)&=\frac{1}{2}\mathrm{erfc}\left(\frac{t-t_{e}}{ \sqrt {\mathstrut 2\sigma _{e}^{2}}}\right)+\frac{e^{-(t-t_{e}')/\tau _{e}}}{2}\mathrm{erfc}\left(-\frac{t-t_{e}'}{\sqrt {\mathstrut 2\sigma _{e}^{2}}}\right),
  \label{eq:fitting-function}
\end{align}
where $t_{e}$ and $\sigma _{e}^{2}$ represent the average and variance of the arrival time of the excited wavepacket at the crossing region, respectively, and $\tau _{e}$ is the de-excitation time constant via the crossing region, $t_{e}'\equiv t_{e}+2\sigma _{e}^{2}/\tau _{e}$, and $\mathrm{erfc}(x)$ is the complementary error function, $\mathrm{erfc}(x)\equiv 1-\mathrm{erf}(x)$.
For the details of the fitting function, see Appendix~\ref{sec:fitting-function}.
Because we have assumed $\alpha _{\mathrm{E}}=1$, almost all population is in the excited state at $t=0$, i.e.~$u_{e}(0)\simeq 1$.
Then the wavepacket moves toward the crossing region and is de-excited through the NAC.
From the fitting parameters, the timescale of the wavepacket motion is estimated as $t_{e}=0.62~\mathrm{ps}$, while the timescale of the non-adiabatic transition is $\tau _{e}=0.41~\mathrm{ps}$.

\begin{figure}
  \centering
  \includegraphics[scale=\SingleColFigScale]{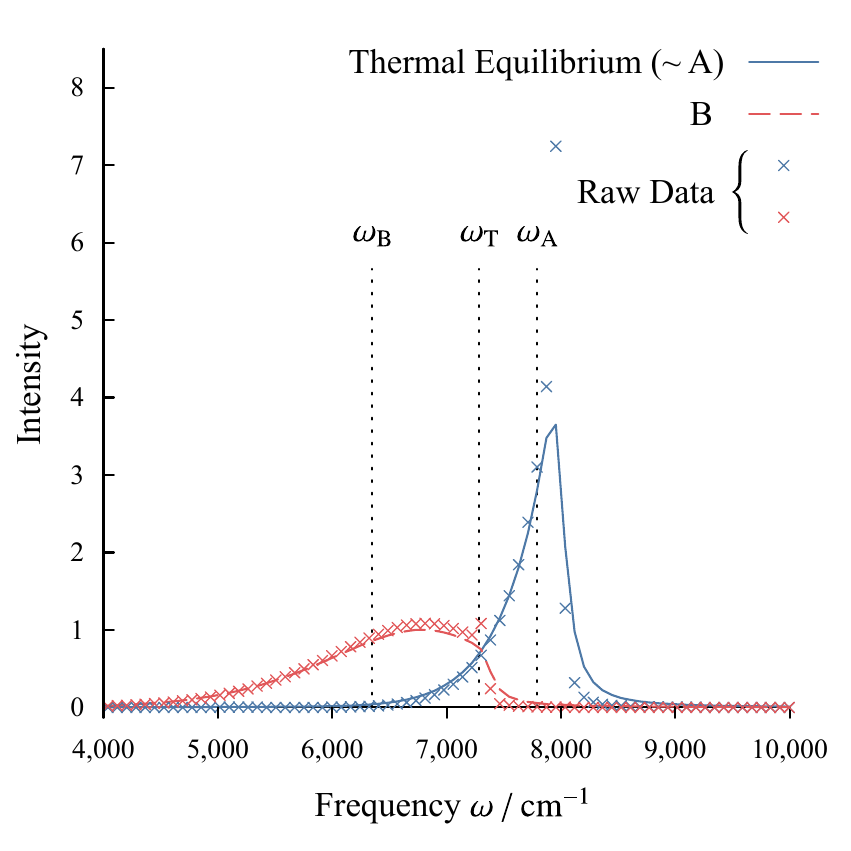}
  \caption{
    Linear absorption spectra from the thermal equilibrium state (the blue solid curve; the wavepacket is almost localized at $\mathrm{A}$) and from the thermal metastable state $\mathrm{B}$ (the red dashed curve).
    The intensity is normalized with respect to the maximum peak of the absorption of $\mathrm{B}$.
    Here, we employed a window function, $e^{-t/\tau }$ with $\tau =80~\mathrm{fs}$, to carry out the Fourier transformations to suppress the cost of numerical calculations:
    The $\times $ symbols represent the results without the exponential window function.
  }
  \label{fig:linear-absorption}
\end{figure}
\begin{figure}
  \centering
  \includegraphics[scale=\SingleColFigScale]{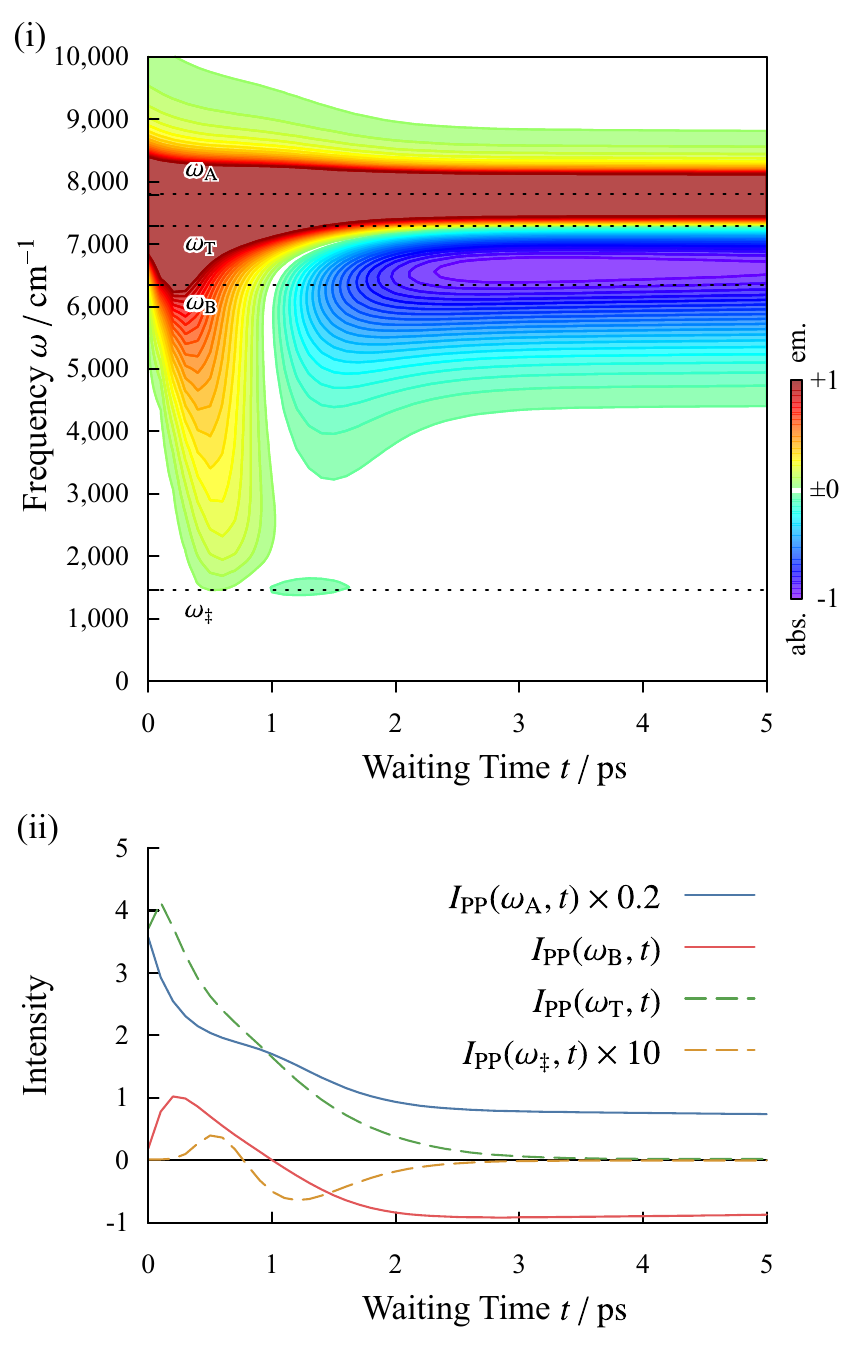}
  \caption{
    Pump-probe spectrum for the photoisomerization process from $\mathrm{A}$ to $\mathrm{B}$.
    (i) The 2D color map of pump-probe spectrum.
    The red and blue areas represent emission and absorption, respectively.
    The intensity is normalized with respect to its maximum of absorption peak at $t\simeq 4~\mathrm{ps}$.
    (ii) The intensities of the spectrum at $\omega =\omega _{\mathrm{A}}$ (blue solid), $\omega _{\mathrm{B}}$ (red solid), $\omega _{\mathrm{T}}\equiv (U_{e}(\theta _{\mathrm{T}})-U_{g}(\theta _{\mathrm{T}}))/\hbar =7,283~\mathrm{cm}^{-1}$ (green dashed), and $\omega _{\ddagger }$ (yellow dashed) are plotted as functions of the waiting time.
    The results of $I_{\mathrm{PP}}(\omega _{\mathrm{A}},t)$ and $I_{\mathrm{PP}}(\omega _{\ddagger },t)$ were multiplied by $0.2$ and $10$ for comparison.
  }
  \label{fig:pump-probe-A}
\end{figure}
Figures~\ref{fig:linear-absorption} and ~\ref{fig:pump-probe-A} present linear absorption and pump-probe spectra from the thermal equilibrium state for the photoisomerization process depicted in Fig.~\ref{fig:nonadiabatic_transition}.
In Figs.~\ref{fig:linear-absorption}, \ref{fig:pump-probe-A} and all of the following results, a window function, $e^{-t/\tau }$ with $\tau =80~\mathrm{fs}$, was employed in Eqs.~\eqref{eq:response-spectrum} and \eqref{eq:r-nr-spectrum} in order to suppress the range of the time integration of Fourier transforms.
In Fig.~\ref{fig:pump-probe-A}(i), the absorption peak centered at the electronic resonant frequency, $\omega _{\mathrm{A}}\equiv (U_{e}(\theta _{\mathrm{A}})-U_{g}(\theta _{\mathrm{A}}))/\hbar =7,792~\mathrm{cm}^{-1}$, is observed at the Frank-Condon point $\theta =\theta _{\mathrm{A}}$.
In pump-probe spectrum in Fig.~\ref{fig:pump-probe-A}, the emission peaks for the ground state bleaching (GSB) and stimulated emission (SE) processes are observed near $\omega =\omega _{\mathrm{A}}$ at $t=0$.
Then, the SE peak moves to a lower frequency region following the wavepacket motion (time-dependent Stokes shift) toward $\omega _{\ddagger }\equiv (U_{e}(\theta _{\ddagger })-U_{g}(\theta _{\ddagger }))/\hbar =1,455~\mathrm{cm}^{-1}$ at $t\simeq t_{e}$.
Here, $\theta _{\ddagger }$ is located at the minimum point of the excited BO PES.
After the non-adiabatic transition, the wavepacket moves into the photo-product state $\mathrm{B}$ and the absorption peak at $\omega _{\mathrm{B}}\equiv (U_{e}(\theta _{\mathrm{B}})-U_{g}(\theta _{\mathrm{B}}))/\hbar =6,345~\mathrm{cm}^{-1}$ is observed as a photoinduced absorption peak.
As this indicates, the information for a single photoisomerization process can be obtained from pump-probe spectroscopy.

The bilinear (linear-linear; LL) coordinate-bath coupling in Eq.~\eqref{eq:bath} gives raise to fluctuation of the coordinate $q$ and this indirectly causes frequency fluctuation of the electronic resonant frequency, i.e.~$(U_{e}(\theta )-U_{g}(\theta ))$, as was demonstrated using the LL Brownian displaced oscillators model \cite{mukamel1999book, tanimura2006jpsj}, which reduces to the stochastic two-level model that describes inhomogeneous broadening in the overdamped limit.
In the present model, however; the effect of this fluctuation is minor and the broadening we observed here mainly arises from the width of the wavepackets that distributes and moves in the ground/excited states.
Hence the broadening has a dynamical nature that should be distinguished from the inhomogeneous broadening.
Therefore, we refer to this phenomenon as the ``dynamical Stokes broadening'', and this causes diagonally elongated peaks in two-dimensional correlation spectra as shown below.
The difference of the peak intensities of the thermal equilibrium state and thermal metastable state in Fig.~\ref{fig:linear-absorption} arises from this broadening.
In the infinite friction limit, $\zeta \rightarrow \infty $ in Eq.~\eqref{eq:ms-lt-qse}, the wavepackets completely settle and this broadening agrees with the inhomogeneous broadening.

\subsection{Thermalization}
\label{sec:thermalization}
Next, we investigate a thermalization process from $\mathrm{B}$ to $\mathrm{A}$.
To estimate the timescale of the thermalization, we employ the classical Boltzmann distribution in the adiabatic ground state as the initial state $f_{gg}(\theta ,t=0)$ in the region $\theta _{\mathrm{P}}\leq q\leq \pi $, and otherwise $\bm{f}(\theta ,t=0)=0$.
Then we integrate Eq.~\eqref{eq:ms-lt-qse} to investigate the time evolution of the system.
We define the population of $\mathrm{A}$ and $\mathrm{B}$ as
\begin{subequations}
  \begin{align}
    u_{\mathrm{A}}(t)&\equiv \int _{0}^{\theta _{\mathrm{P}}}\!d\theta \,f_{gg}(\theta ,t)\\
    \intertext{and}
    u_{\mathrm{B}}(t)&\equiv \int _{\theta _{\mathrm{P}}}^{\pi }\!d\theta \,f_{gg}(\theta ,t),
  \end{align}
\end{subequations}
respectively.

\begin{figure}
  \centering
  \includegraphics[scale=\SingleColFigScale]{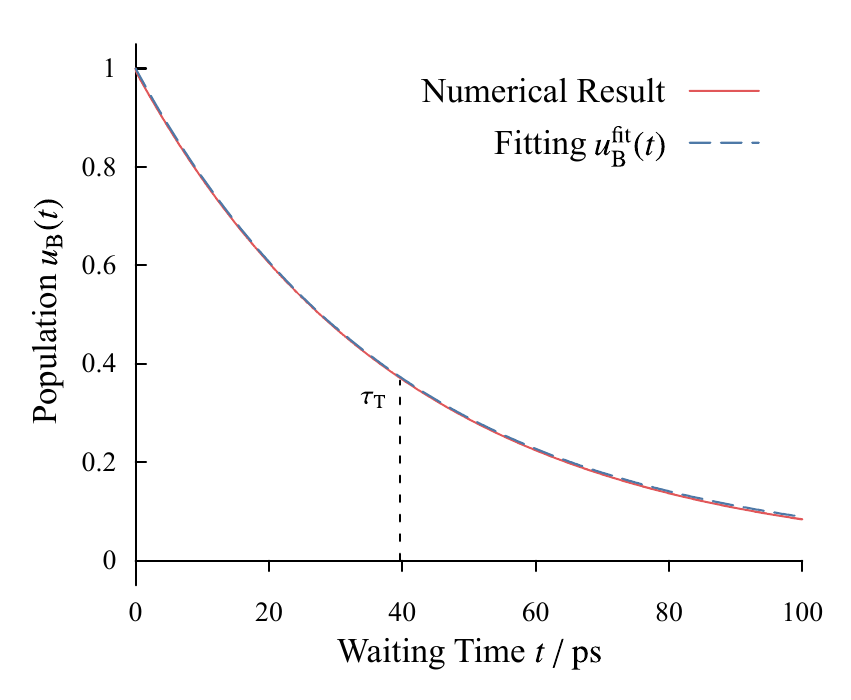}
  \caption{
    The population of $\mathrm{B}$ as a function of time, $u_{\mathrm{B}}(t)$.
    The solid curve represents the numerical result, whereas the dashed curve represents the fitted result using Eq.~\eqref{eq:u_B-fit}.
    The fitting parameters are $\tau _{\mathrm{T}}=39.6~\mathrm{ps}$ and $\alpha _{\mathrm{T}}^{+}=0.99$.
  }
  \label{fig:thermalization}
\end{figure}
In Fig.~\ref{fig:thermalization}, we display the calculated result and the fitted result using the fitting function
\begin{align}
  u_{\mathrm{B}}^{\mathrm{fit}}(t)&=1-\alpha _{\mathrm{T}}^{+}+\alpha _{\mathrm{T}}^{+}e^{-t/\tau _{\mathrm{T}}},
  \label{eq:u_B-fit}
\end{align}
where $\tau _{\mathrm{T}}$ and $\alpha _{\mathrm{T}}^{+}$ are the fitting parameters.
These parameters relate with the rate equation of the thermalization process expressed as
\begin{align}
  \left\{
  \begin{aligned}
    \frac{du_{\mathrm{A}}}{dt}&=-k_{\mathrm{A}'\rightarrow \mathrm{B}}u_{\mathrm{A}}+k_{\mathrm{B}\rightarrow \mathrm{A}'}u_{\mathrm{B}}\\
    \frac{du_{\mathrm{B}}}{dt}&=-k_{\mathrm{B}\rightarrow \mathrm{A}'}u_{\mathrm{B}}+k_{\mathrm{A}'\rightarrow \mathrm{B}}u_{\mathrm{A}},
  \end{aligned}\right.
  \label{eq:thermal-rate}
\end{align}
where $k_{\mathrm{\alpha}\rightarrow \mathrm{\beta}}$ is the rate constant for the process $\alpha \rightarrow \beta $ under the initial condition $u_{\mathrm{B}}(0)=1$.
The fitting parameters in Eq. \eqref{eq:u_B-fit} are then determined as $\tau _{\mathrm{T}}\equiv 1/(k_{\mathrm{A}'\rightarrow \mathrm{B}}+k_{\mathrm{B}\rightarrow \mathrm{A}'})$ and $\alpha _{\mathrm{T}}^{+}\equiv k_{\mathrm{B}\rightarrow \mathrm{A}'}/(k_{\mathrm{A}'\rightarrow \mathrm{B}}+k_{\mathrm{B}\rightarrow \mathrm{A}'})$, which describe the time decay constant and ratio of the forward conversion ($\mathrm{B}\rightarrow \mathrm{A}'$) in the thermalization, respectively.

The fitted result indicates that the timescale of the thermalization is $\tau _{\mathrm{T}}=39.6~\mathrm{ps}$, which is approximately $100$ times slower than the time scales of photoisomerization $t_{e}$ and $\tau _{e}$.
Thus, the rate-limiting process of the molecular motor system described by the present model is the thermalization.
Note that, as shown in Sec.~\ref{sec:exam-qltc}, while the QLT correction terms of the MS-LT-QSE is important to obtain physically accurate non-adiabatic transition process, these terms do not play a role in the thermalization process, because the temperature is high enough in comparison with the characteristic vibrational frequency near $\mathrm{B}$, $1/\beta \hbar =208~\mathrm{cm}^{-1}>\Omega _{\mathrm{B}}=79.7~\mathrm{cm}^{-1}$.
Thus the reaction velocity estimated from the classical Smoluchowski equation~\eqref{eq:se}, $1/k_{\mathrm{K}}=38.8~\mathrm{ps}$, where $k_{\mathrm{K}}\equiv \Omega _{\mathrm{B}}\Omega _{\mathrm{T}}e^{-\beta \Delta E_{\mathrm{T}}}/2\pi \zeta $ \cite{kramers1940p} is closer to $\tau _{\mathrm{T}}$.

\begin{figure}
  \centering
  \includegraphics[scale=\SingleColFigScale]{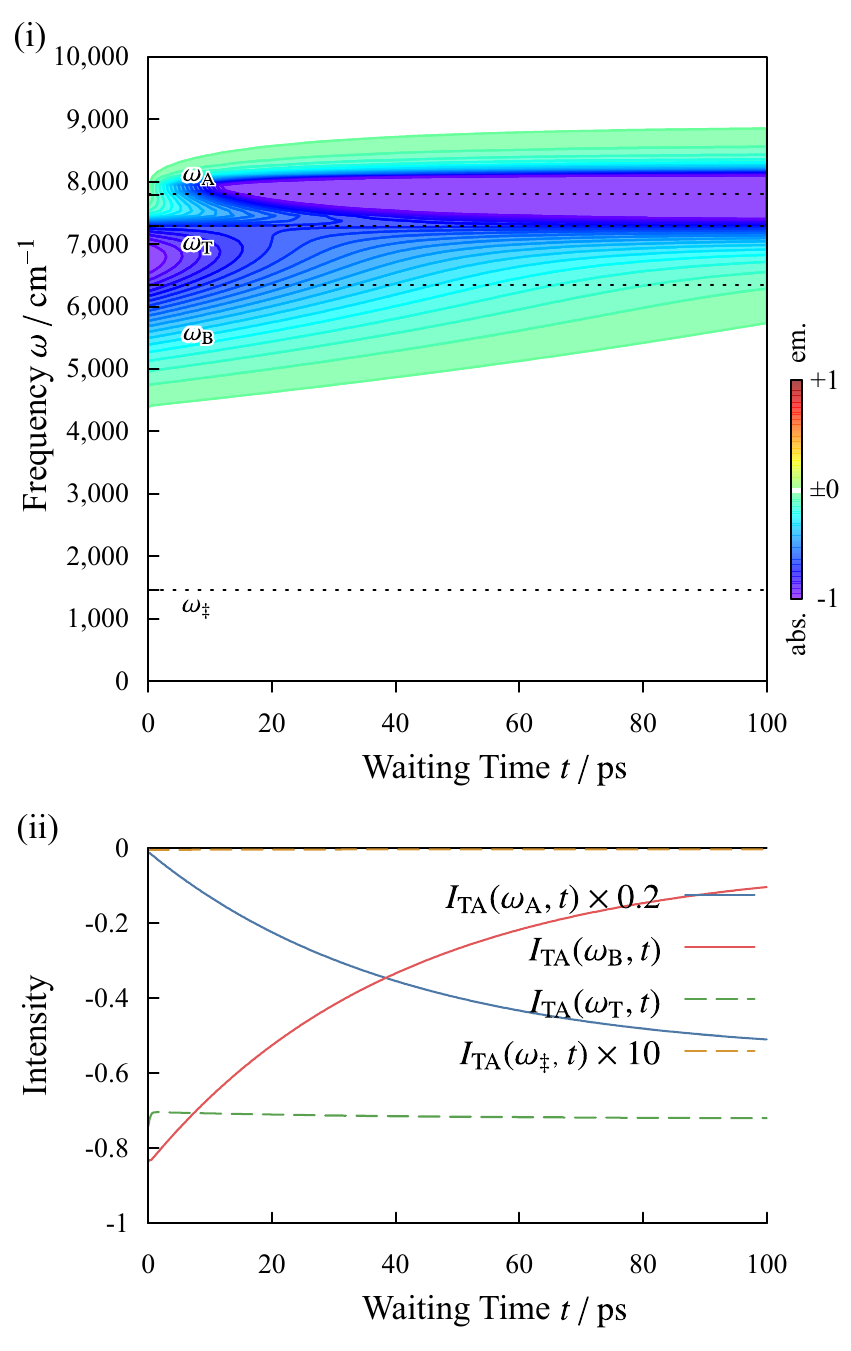}
  \caption{
    Transient absorption (TA) spectrum for the thermalization process from $\mathrm{B}$ to $\mathrm{A}$.
    (i) The 2D color map of the TA spectrum.
    The red and blue areas represent emission and absorption, respectively.
    The intensity is normalized with respect to the maximum peak at $t=0~\mathrm{fs}$.
    (ii) The intensities of the spectrum at $\omega =\omega _{\mathrm{A}}$ (blue solid), $\omega _{\mathrm{B}}$ (red solid), $\omega _{\mathrm{T}}$ (green dashed), and $\omega _{\ddagger }$ (yellow dashed) are plotted as functions of the waiting time.
    The results of $I_{\mathrm{TA}}(\omega _{\mathrm{A}},t)$ and $I_{\mathrm{TA}}(\omega _{\ddagger },t)$ were multiplied by $0.2$ and $10$ for comparison.
  }
  \label{fig:transient-absorption-B}
\end{figure}
Figure~\ref{fig:transient-absorption-B} depicts the transient absorption (TA) spectrum for the thermalization process analyzed in Fig.~\ref{fig:thermalization}.
Here, the absorption spectrum at $t=0$ is illustrated in Fig.~\ref{fig:linear-absorption}.
Following the thermalization of the wavepacket from $\mathrm{B}$ to $\mathrm{A}$, the absorption peak centered at $\omega =\omega _{\mathrm{B}}$ vanishes, while that centered at $\omega =\omega _{\mathrm{A}}$ appears.
As was also shown in experiments \cite{klok2008jacs}, TA spectroscopy has the capability to investigate the thermalization process.

\subsection{Stationary rotating process driven by pulse repetition}
\label{sec:pulse-repetition}
Up to now, we treated the photoisomerization and thermalization processes separately.
Here we study the stationary rotating process driven by periodical pulses characterized by the average rotational speed.
For this purpose, we simulate the time-evolution of the system under $N$ periodic pulses with interval $\tau _{\mathrm{p}}$.
The wavepacket is then expressed as
\begin{align}
  \bm{f}_{\tau _{\mathrm{p}}}^{N}(\theta ,\Delta t)=\mathcal{G}(\Delta t)\mathcal{G}_{\mathrm{p}}(\mathcal{G}(\tau _{\mathrm{p}})\mathcal{G}_{\mathrm{p}})^{N-1}\bm{f}(\theta ,0),
  \label{eq:n-perturbed-distribution}
\end{align}
where $\mathcal{G}$ is Green's function obtained by integrating Eq.~\eqref{eq:ms-lt-qse}, $\tau _{\mathrm{p}}$ is the pulse repetition interval, and $\Delta t$ is the elapsed time after the last pulse interaction.
The flux $j_{\tau _{\mathrm{p}}} (\theta ,t)$ for the above process is evaluated from Eq.~\eqref{eq:flux}.
Then we further introduce the accumulated flux, $J_{\tau _{\mathrm{p}}}^{N}(\theta ,\Delta t)$, and averaged flux, $\bar{\jmath }_{\tau _{\mathrm{p}}}^{N}(\theta ,\Delta t)$, after the $N$th pulse excitation defined using Eq.~\eqref{eq:n-perturbed-distribution} as
\begin{align}
  J_{\tau _{\mathrm{p}}}^{N}(\theta ,\Delta t)&=\int _{0}^{\Delta t}\!ds\,j_{\tau _{\mathrm{p}}}(\theta ,s)\\
  \intertext{and}
  \bar{\jmath }_{\tau _{\mathrm{p}}}^{N}(\theta ,\Delta t)&=\frac{J_{\tau _{\mathrm{p}}}^{N}(\theta ,\Delta t)}{\Delta t},
\end{align}
respectively.
\begin{figure}
  \centering
  \includegraphics[scale=\SingleColFigScale]{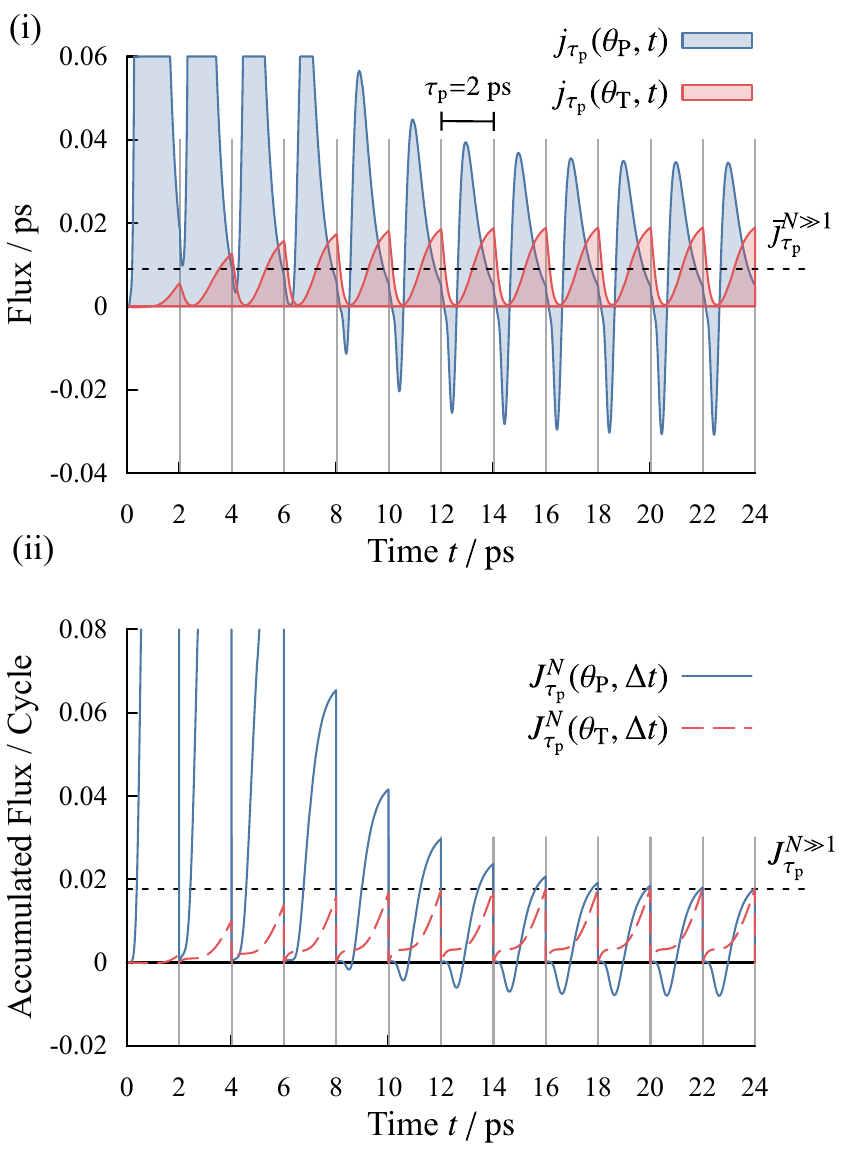}
  \caption{
    (i) The flux at the barrier tops of the photoisomerization, $j_{\tau _{\mathrm{p}}}(\theta _{\mathrm{P}},t)$, (the blue curve) and the flux at the barrier tops of the thermalization, $j_{\tau _{\mathrm{p}}}(\theta _{\mathrm{T}},t)$, (the red curves) are depicted as functions of the elapsed time under the pulse repetition interval $\tau _{\mathrm{p}}=2~\mathrm{fs}$.
    After $t=24~\mathrm{ps}$, the wavepacket movement reaches the time-dependent stationary-state, and the flux also changes periodically.
    The dotted line is average value of the stationary solution, $\bar{\jmath }_{\tau _{\mathrm{p}}}^{N\gg 1}$.
    (ii) Accumulated flux per pulse cycle for photoisomerization, $J_{\tau _{\mathrm{p}}}^{N}( \theta _{\mathrm{P}},\Delta t)$ (the blue curve) and for thermalization $J_{\tau _{\mathrm{p}}}^{N}(\theta _{\mathrm{T}},\Delta t)$ (the red curve) obtained from the results in (i).
    These results correspond the areas of the curves per cycle depicted in (i).
    The dotted line is that of the stationary solution, $J_{\tau _{\mathrm{p}}}^{N\gg 1}$.
  }
  \label{fig:stationary-flow}
\end{figure}
After apply sufficiently many pulses ($N\gg 1$), the distribution changes periodically in time (i.e.~$\bm{f}_{\tau _{\mathrm{p}}}^{N+1}(\theta ;\Delta t)\simeq \bm{f}_{\tau _{\mathrm{p}}}^{N}(\theta ,\Delta t)$).
Accordingly, the flux also changes periodically (e.g.~$J_{\tau _{\mathrm{p}}}^{N+1}(\theta ;\Delta t)\simeq J_{\tau _{\mathrm{p}}}^{N}(\theta ,\Delta t)$) as illustrated in Fig.~\ref{fig:stationary-flow}.
Note that, while the results in the range $0~\mathrm{ps}\leq t\leq 24~\mathrm{ps}$ were displayed in Fig.~\ref{fig:stationary-flow}, the numerical evaluation of $J_{\tau _{\mathrm{p}}}^{N\gg 1}$ and $\bar{\jmath }_{\tau _{\mathrm{p}}}^{N\gg 1}$ were performed for $t\gtrsim 500~\mathrm{ps}$.

We next investigate the performance of the motor as a function of the pulse repetition rate $\phi \equiv 1/\tau _{\mathrm{p}}$.
For this purpose, we introduce the average rotational speed $\eta (\phi )$ as follows.
The flux $\bar{\jmath }_{\tau _{\mathrm{p}}}^{N}(\theta ,\tau _{\mathrm{p}})$ represents the averaged value of the flow of population at $\theta $ during the $N$th pulse cycle, which is position independent under the stationary condition (i.e.~$\bar{\jmath }_{\tau _{\mathrm{p}}}^{N}=\bar{\jmath }_{\tau _{\mathrm{p}}}^{N}(\theta ,\tau _{\mathrm{p}})$).
Because the population is normalized for the half rotational motion $0\leq \theta <\pi $ (see Fig.~\ref{fig:PES}), the average time period for the half rotational motion is expressed as $1/\bar{\jmath }_{\tau _{\mathrm{p}}}^{N\gg 1}$.
Hence, the rotational speed (i.e.~the number of rotations per unit time) is described by $\eta (\phi )\equiv 0.5\,\bar{\jmath }_{\tau _{\mathrm{p}}}^{N\gg 1}$ ($[\mathrm{T}^{-1}]$).
The average time duration of a single rotation is then given by $1/\eta (\phi )$.

We then estimate the average rotational speed using the fact that the time scales of the photoisomerization and thermalization are very different.
When we ignore the details of the photoisomerization process, we can roughly estimate its flux as the following:
Consider the case that the population of $\mathrm{A}$ and $\mathrm{B}$ are given by $u_{\mathrm{A}}^{(N)}$ and $u_{\mathrm{B}}^{(N)}$, where $u_{\mathrm{A}}^{(N)}+u_{\mathrm{B}}^{(N)}=1$.
After the $(N+1)$th pulse is applied, a part of the population of $\mathrm{A}$, $\alpha _{\mathrm{E}}y_{\mathrm{P}}^{+}u_{\mathrm{A}}^{(N)}$, is converted to $\mathrm{B}$, where $y_{\mathrm{P}}^{+}$ is the yield of the product state in the photoisomerization process $\mathrm{A}\rightarrow \mathrm{B}$.
At the same time, a part of the population of $\mathrm{B}$, $\alpha _{\mathrm{E}}y_{\mathrm{P}}^{-}u_{\mathrm{B}}^{(N)}$, is converted to $\mathrm{A}$ due to the backward photoisomerization, where $y_{\mathrm{P}}^{-}$ is the yield of the backward photoisomerization $\mathrm{B}\rightarrow \mathrm{A}$.
By using the effective yields of the photoisomerization that include the photoexcitation process, $\bar{y}_{\mathrm{p}}^{+}\equiv \alpha _{\mathrm{E}}y_{\mathrm{P}}^{+}$ and $\bar{y}_{\mathrm{p}}^{-}\equiv \alpha _{\mathrm{E}}y_{\mathrm{P}}^{-}$, the populations of $\mathrm{A}$ and $\mathrm{B}$ can be described by $(1-\bar{y}_{\mathrm{P}}^{+})u_{\mathrm{A}}^{(N)}+\bar{y}_{\mathrm{P}}^{-}u_{\mathrm{B}}^{(N)}$ and $\bar{y}_{\mathrm{P}}^{+}u_{\mathrm{A}}^{(N)}+(1-\bar{y}_{\mathrm{P}}^{-})u_{\mathrm{B}}^{(N)}$, respectively.
By solving Eq.~\eqref{eq:thermal-rate} for $\Delta t=\tau _{\mathrm{p}}$, the population of $\mathrm{A}'$ ($=\mathrm{A}$) is evaluated as
\begin{align}
  \begin{split}
    u_{\mathrm{A}}^{(N+1)}&=[1-(\bar{y}_{\mathrm{P}}^{+}+\bar{y}_{\mathrm{P}}^{-})]e^{-\tau _{\mathrm{p}}/\tau _{\mathrm{T}}}u_{\mathrm{A}}^{(N)}\\
    &\quad +\bar{y}_{\mathrm{P}}^{-}e^{-\tau _{\mathrm{p}}/\tau _{\mathrm{T}}}+\alpha _{\mathrm{T}}^{+}(1-e^{-\tau _{\mathrm{p}}/\tau _{\mathrm{T}}}).
  \end{split}
\end{align}
For the stationary current, $u_{\mathrm{A}}^{(N+1)}=u_{\mathrm{A}}^{(N)}\equiv u_{\mathrm{A}}^{(\infty )}$, the flux is given by $\bar{y}_{\mathrm{P}}^{+}u_{\mathrm{A}}^{(\infty )}-\bar{y}_{\mathrm{P}}^{-}u_{\mathrm{B}}^{(\infty )}$.

Then the average rotational speed is evaluated as
\begin{align}
  \eta ^{\mathrm{IPI}}(\phi )=\eta _{+}^{\mathrm{IPI}}(\phi )-\eta _{-}^{\mathrm{IPI}}(\phi ),
  \label{eq:ipi-approximation}
\end{align}
where the positive (+) and negative (-) currents in the stationary rotating process are expressed as 
\begin{align}
  \eta _{\pm }^{\mathrm{IPI}}(\phi )&=\frac{0.5\,\alpha _{\mathrm{T}}^{\pm }\bar{y}_{\mathrm{P}}^{\pm }\phi }{1+(\bar{y}_{\mathrm{P}}^{+}+\bar{y}_{\mathrm{P}}^{-})/(e^{1/\tau _{\mathrm{T}}\phi }-1)}
  \label{eq:ipi-approximation-2}
\end{align}
and $\alpha _{\mathrm{T}}^{-}\equiv 1-\alpha _{\mathrm{T}}^{+}=k_{\mathrm{A}'\rightarrow \mathrm{B}}/(k_{\mathrm{A}'\rightarrow \mathrm{B}}+k_{\mathrm{B}\rightarrow \mathrm{A}'})$ is the backward conversion ratio ($\mathrm{A}'\rightarrow \mathrm{B}$) in the thermalization process.
%% \modified{\deleted{
%% The slip under the IPI approximation is evaluated as
%% \begin{align}
%%   s^{\mathrm{IPI}}(\phi )&=1-2(\eta _{+}^{\mathrm{IPI}}(\phi )-\eta _{-}^{\mathrm{IPI}}(\phi )).
%%   \label{eq:ipi-approximation-s}
%% \end{align}
%% }}

Because the denominator of Eq.~\eqref{eq:ipi-approximation-2} is always positive, the inequality relation of the positive numerator,
\begin{align}
  \alpha _{\mathrm{T}}^{+}\bar{y}_{\mathrm{P}}^{+}&>\alpha _{\mathrm{T}}^{-}\bar{y}_{\mathrm{P}}^{-},
  \label{eq:positive-rotation}
\end{align}
represents the condition for the positive rotation.
This indicates that, even if the photoisomerization or thermalization exhibits a negative tendency (i.e.~ $\bar{y}_{\mathrm{P}}^{+}<\bar{y}_{\mathrm{P}}^{-}$ or $\alpha _{\mathrm{T}}^{+}<\alpha _{\mathrm{T}}^{-}$), the condition Eq.~\eqref{eq:positive-rotation} guarantees a positive stationary rotation.
Under this condition, $\eta ^{\mathrm{IPI}}(\phi )$ becomes positive, and increases monotonically with the increase of $\phi $, i.e.~$\partial \eta ^{\mathrm{IPI}}(\phi )/\partial \phi >0$.

\begin{figure}
  \centering
  \includegraphics[scale=\SingleColFigScale]{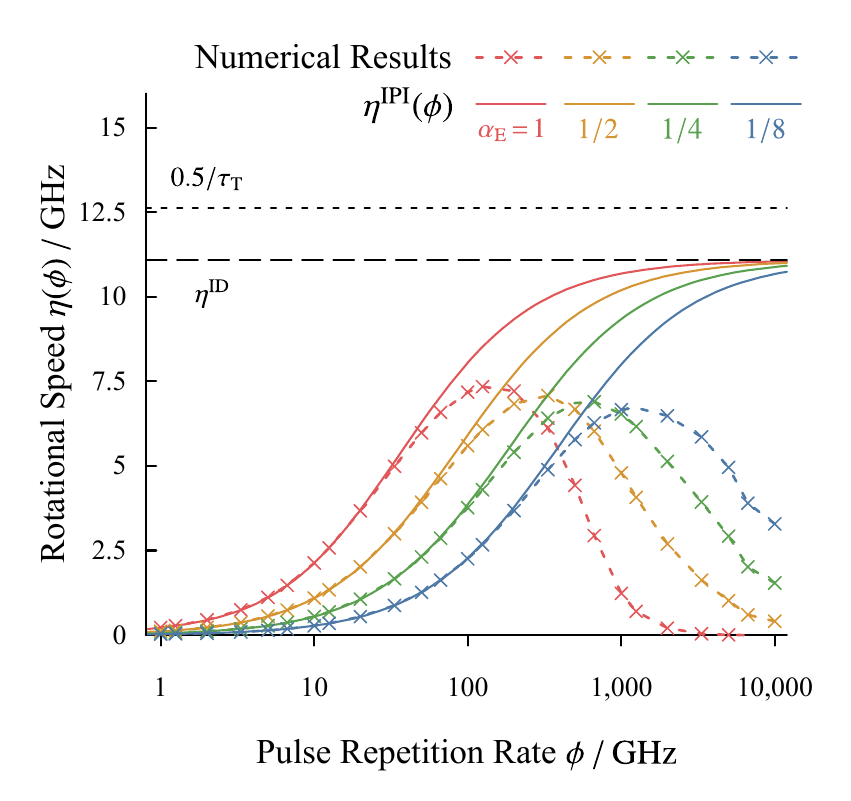}
  \caption{
    The average rotational speed under pulse repetition driving, $\eta (\phi )$, as a function of the pulse repetition rate, $\phi $, for various values of $\alpha _{\mathrm{E}}$.
    The $\times $ symbols represent the numerical results, while the solid curves represent the approximated results using Eq.~\eqref{eq:ipi-approximation}.
    The dashed line represents the maximum rotational speed under the IPI approximation, $\eta ^{\mathrm{ID}}$, while the dotted line represents the rotational speed of the thermalization, $0.5/\tau _{\mathrm{T}}$.
    Note that $1~\mathrm{GHz}=1~\mathrm{ns}^{-1}=6\times 10^{10}~\mathrm{rpm}$, where $\mathrm{rpm}$ stands for the revolutions/rotations per minute.
    %% \modified{\deleted{(ii) The slip, $s(\phi )$, as a function of $\phi $.
    %% The red $\times $ symbols represent the numerical results, while the blue solid curve represent the approximated results using Eq.~\eqref{eq:ipi-approximation-s}.}}
  }
  \label{fig:speed}
\end{figure}
In Fig.~\ref{fig:speed}, we display the calculated results of the average rotational speed $\eta (\phi )$ as a function of the pulse repetition rate, $\phi $, for various values of $\alpha _{\mathrm{E}}$ that were determined  from $\varphi _{\mathrm{p}}$ in Eq.~\eqref{eq:photoexcitation} in the range of $0\leq \varphi _{\mathrm{p}}\leq \pi /2$.
To adopt Eq.~\eqref{eq:ipi-approximation-2}, we set $\tau _{\mathrm{T}}$ and $\alpha _{\mathrm{T}}^{+}$ obtained in Sec.~\ref{sec:thermalization}.
Then the values of the yields were evaluated from the numerical calculations of single photoisomerization processes as $y_{\mathrm{P}}^{+}=0.45$ and $y_{\mathrm{P}}^{-}=0.057$ (see Appendix ~\ref{sec:yields}).
Note that, when we ignore the thermalization process, the equilibrium populations of $\mathrm{A}$ and $\mathrm{B}$ under light can be approximated by $y_{\mathrm{P}}^{-}/(y_{\mathrm{P}}^{+}+y_{\mathrm{P}}^{-})$ and $y_{\mathrm{P}}^{+}/(y_{\mathrm{P}}^{+}+y_{\mathrm{P}}^{-})$.
In the present model, thus we have $u_{\mathrm{A}}:u_{\mathrm{B}}=11:89$.

The maximum (ideal) rotational speed is achieved in the fast pulse repetition limit, $\tau _{\mathrm{T}}\phi \gg 1$, as
\begin{align}
  \eta ^{\mathrm{IPI}}(\phi )&\mathop{\rightarrow }_{\phi \rightarrow \infty }\eta ^{\mathrm{ID}}\equiv \frac{0.5}{\tau _{\mathrm{T}}}\frac{\alpha _{\mathrm{T}}^{+}\bar{y}_{\mathrm{P}}^{+}-\alpha _{\mathrm{T}}^{-}\bar{y}_{\mathrm{P}}^{-}}{\bar{y}_{\mathrm{P}}^{+}+\bar{y}_{\mathrm{P}}^{-}}&(\leq 0.5/\tau _{\mathrm{T}})
  \label{eq:ideal-speed}.
\end{align}
Thus, the upper-limit of $\eta (\phi )$ is determined by the timescale of the thermalization, $0.5/\tau _{\mathrm{T}}$.
The higher pulse repetition rate produces the higher rotational speed in the IPI approximation, $\eta ^{\mathrm{IPI}}(\phi )$.
In actual cases, however, the average rotational speed $\eta (\phi )$ has a maximum as a function of $\phi $:
When the pulse repetition rate is too large, laser interactions cause not only excitation but also stimulated de-excitation among the electronic states.
This de-excitation process suppresses the efficiency of the photoisomerization processes.
In this regime, the details of the fast photoisomerization process must be accounted for to attain the maximum rotational speed.
The effective time interval of the laser interaction (e.g.~average time interval $\tau _{\mathrm{p}}/\alpha _{\mathrm{E}}$) increases with decrease of $\alpha _{\mathrm{E}}$.
Because the stimulated de-excitation decreases with decrease of $\alpha _{\mathrm{E}}$ even when $\phi $ is large, the peak position of $\eta (\phi )$ shifts to the high-frequency region for small $\alpha _{\mathrm{E}}$.
%% \begin{align}
%%   \lim _{N\rightarrow \infty }\sum _{k}^{N}\binom{N}{k}\frac{N\tau _{\mathrm{p}}}{k+1}\alpha _{E}^{k}(1-\alpha _{E})^{N-k}&=\lim _{N\rightarrow \infty }\frac{N}{N+1}\frac{\tau _{\mathrm{p}}}{\alpha _{E}}\sum _{k}^{N}\binom{N+1}{k+1}\alpha _{E}^{k+1}(1-\alpha _{E})^{N+1-(k+1)}\\
%%   &=\lim _{N\rightarrow \infty }\frac{N}{N+1}\frac{\tau _{\mathrm{p}}}{\alpha _{E}}\left(\sum _{k=0}^{N+1}\binom{N+1}{k}\alpha _{E}^{k}(1-\alpha _{E})^{N-k}-(1-\alpha _{E})^{N}\right)\\
%%   &=\frac{\tau _{\mathrm{p}}}{\alpha _{E}}
%% \end{align}

%% \modified{\deleted{
%% In Fig.~\ref{fig:speed}(ii), we display the calculated results of the slip $s(\phi )$ as a function of $\phi $.
%% The minimum (ideal) slip is achieved in the slow pulse repetition limit, $\tau _{\mathrm{T}}\phi \ll 1$, as
%% \begin{align}
%%   s^{\mathrm{IPI}}(\phi )&\mathop{\rightarrow }_{\phi \rightarrow +0}s^{\mathrm{ID}}(\phi )\equiv 1-\alpha _{\mathrm{E}}(\alpha _{\mathrm{T}}^{+}\bar{y}_{\mathrm{P}}^{+}-\alpha _{\mathrm{T}}\bar{y}_{\mathrm{P}}^{-}).
%%   \label{eq:ideal-s}
%% \end{align}
%% Equation~\eqref{eq:ideal-s} indicates that, the synchronous rotation (i.e.~$s^{\mathrm{ID}}(\phi )=0$) cannot be achieved, except the case $\bar{y}_{\mathrm{P}}^{+}=1$ and $\alpha _{\mathrm{T}}^{+}=1$, because inefficient photoisomerization or thermalization processes hinder the complete synchronous rotation.
%% }}

\begin{figure*}
  \centering
  \includegraphics[scale=\BigDoubleColFigScale]{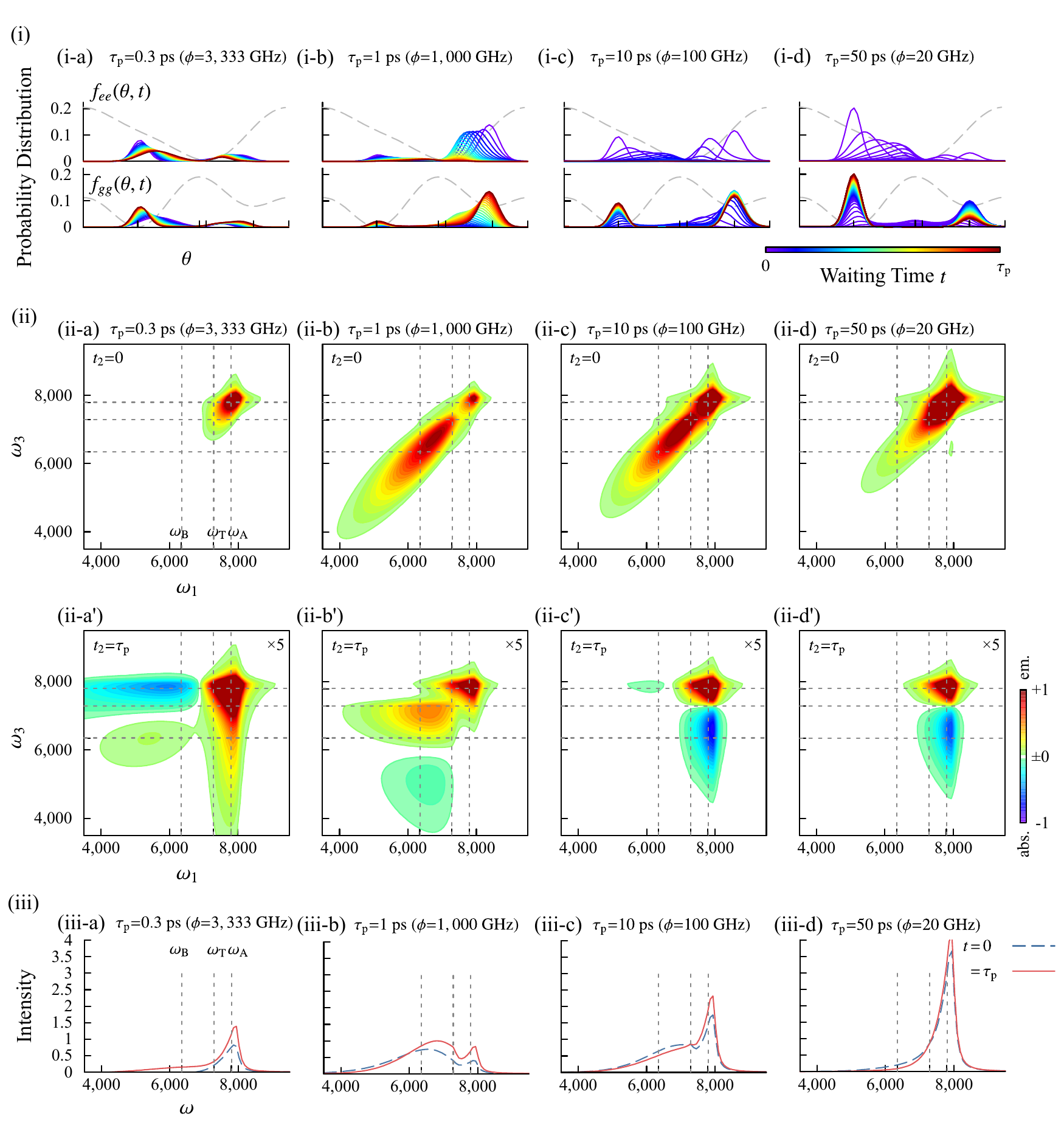}
  \caption{
    (i) Snapshots of the wavepackets in the adiabatic ground/excited states for the pulse repetition intervals, (a) $\tau _{\mathrm{p}}=0.3$, (b) $1$, (c) $10$, and (d) $50~\mathrm{ps}$.
    The colors of the curves represent the different waiting time from $t=0$ to $t=\tau _{\mathrm{p}}$.
    (ii) Two-dimensional correlation spectra, $I_{\mathrm{2D}}(\omega _{3},t_{2},\omega _{1})$, of the stationary rotating process.
    The upper and lower panels represent the spectra for (a)--(d) $t_{2}=0$ and for (a')--(d') $t_{2}=\tau _{\mathrm{p}}$, respectively.
    The red and blue areas represent the emission and absorption, respectively.
    The intensities are normalized with respect to the maximum of the emission peak near $(\omega _{1},\omega _{3})=(\omega _{\mathrm{B}},\omega _{\mathrm{B}})$ in (ii-b), while the intensities of the lower panels are multiplied by $5$.
    (iii) Transient absorption spectra, $I_{\mathrm{TA}}(\omega ,t)$, of the stationary rotating process.
    The dashed blue curves represent $t=0$, whereas the solid red curves represent $t=\tau _{\mathrm{p}}$.
    The intensities are normalized with respect to the maximum of the absorption peak near $\omega =\omega _{\mathrm{B}}$ for $t=0$ in (iii-b).
  }
  \label{fig:2DES}
\end{figure*}
Finally, in order to explore the possibility to characterize the stationary rotating process by means of laser spectroscopy, we calculated 2D correlation spectra and TA spectra for $\hat{\rho }_{\mathrm{tot}}(0)\rightarrow \bm{f}^{N\gg 1}(\theta ,\tau _{\mathrm{p}})$ in the case of $\alpha _{E}=1$.
While 2D correlation spectra in Eqs.~\eqref{eq:r-response} and \eqref{eq:nr-response} that involve the pump excitation can be used to characterize the stimulated processes mainly arising from photoisomerization, TA spectra in Eq.~\eqref{eq:response} that do not involve the excitation are useful to investigate the spontaneous processes mainly arising from thermalization.

Figure~\ref{fig:2DES} displays the (i) snapshots of the wavepacket dynamics, (ii) 2D correlation spectra, and (iii) TA spectra for various values of the pulse repetition interval.
For the fast pulse repetition case, (a) $\tau _{\mathrm{p}}=0.3~\mathrm{ps}$, the distribution is almost localized near $\theta _{\mathrm{A}}$ and $\theta _{\mathrm{B}}$ during whole process (Fig.~\ref{fig:2DES}(i-a)), because the stimulated de-excitation processes occur due to the fast successive pulses before the forward and backward photoisomerization processes are completed.
Therefore, in Fig.~\ref{fig:2DES}(ii-a), the two localized peaks are observed near $(\omega _{1},\omega _{3})=(\omega _{\mathrm{A}},\omega _{\mathrm{A}})$ and $(\omega _{1},\omega _{3})=(\omega _{\mathrm{B}},\omega _{\mathrm{B}})$, which correspond to GSB/SE processes from $\mathrm{A}$ and $\mathrm{B}$, respectively.
The peak from $\mathrm{B}$ is elongated in the $\omega _{1}=\omega _{3}$ direction, because of the dynamical Stokes broadening as explained in Sec.~\ref{sec:photoisomerization}.
In Fig.~\ref{fig:2DES}(ii-a'), the elongated emission peak from $(\omega _{1},\omega _{3})=(\omega _{\mathrm{A}},\omega _{\mathrm{A}})$ to $(\omega _{1},\omega _{3})=(\omega _{\mathrm{A}},\omega _{\ddagger })$ is observed due to the time-dependent Stokes shift as in Fig.~\ref{fig:pump-probe-A}:
This is because the fast pulse driving suppresses the photoisomerization and the wavepacket does not reach the crossing region at $\tau _{\mathrm{p}}$.
We also observe the GSB/SE signals of the backward photoisomerization and absorption from $\mathrm{B}$ as positive and negative peaks near $(\omega _{1},\omega _{3})=(\omega _{\mathrm{B}},\omega _{\mathrm{B}})$ in Fig.~\ref{fig:2DES}(ii-a'), because $t_{2}=\tau _{\mathrm{p}}$ is very short.

For the pulse repetition interval (b) $\tau _{\mathrm{p}}=1~\mathrm{ps}$, the wavepacket is almost localized at $\mathrm{B}$ (Fig.~\ref{fig:2DES}(i-b)), because the fast pulse repetition inhibited the thermalization $\mathrm{B}\rightarrow \mathrm{A}'$ due to the photoexcitation of $\mathrm{B}$, while this allowed the photoisomerization $\mathrm{A}\rightarrow \mathrm{B}$.
Therefore, the peaks near $(\omega _{1},\omega _{3})=(\omega _{\mathrm{B}},\omega _{\mathrm{B}})$ in Fig.~\ref{fig:2DES}(ii-b) and $\omega =\omega _{\mathrm{B}}$ in Fig.~\ref{fig:2DES}(iii-b) at $t=0$ become prominent.
In Fig.~\ref{fig:2DES}(ii-b'), the elongated peak from $(\omega _{1},\omega _{3})=(\omega _{\mathrm{A}},\omega _{\mathrm{A}})$ to $(\omega _{1},\omega _{3})=(\omega _{\mathrm{A}},\omega _{\ddagger })$ is not observed, because the excited wavepacket partially has reached the crossing region.
Because the photoexcitation of $\mathrm{B}$ hinders the thermalization $\mathrm{B}\rightarrow \mathrm{A}'$, the bleaching of the thermalization process is observed as the emission peak near $(\omega _{1},\omega _{3})=(\omega _{\mathrm{B}},\omega _{\mathrm{A}})$, which overlap to the edge of the bleaching peak near $(\omega _{1},\omega _{3})=(\omega _{\mathrm{A}},\omega _{\mathrm{A}})$.

For the pulse repetition interval (c) $\tau _{\mathrm{p}}=10~\mathrm{ps}$, the peaks near $\omega _{1}=\omega _{\mathrm{A}}$ in Fig.~\ref{fig:2DES}(ii-c) and $\omega =\omega _{\mathrm{A}}$ in Fig.~\ref{fig:2DES}(iii-c) at $t=0$ become higher than those of Figs.~\ref{fig:2DES}(ii-b) and \ref{fig:2DES}(iii-b).
This is because the thermalization $\mathrm{B}\rightarrow \mathrm{A'}$ has proceeded during the longer pulse interval.
The negative peak at $(\omega _{1},\omega _{3})=(\omega _{\mathrm{A}},\omega _{\mathrm{B}})$ in Fig.~\ref{fig:2DES}(ii-c') represents the absorption of $\mathrm{B}$ during the photoisomerization $\mathrm{A}\rightarrow \mathrm{B}$.
This profile indicates that the photoisomerization was almost completed during the pulse interval.
Therefore, the average rotating speed for $\tau _{\mathrm{p}}=10~\mathrm{ps}$ (i.e.~$\phi =100~\mathrm{GHz}$) was well described by the IPI approximation, as illustrated in Fig.~\ref{fig:speed}.
The peaks near $(\omega _{1},\omega _{3})=(\omega _{\mathrm{B}},\omega _{\mathrm{B}})$ in Fig.~\ref{fig:2DES}(ii-c') vanish because of the recovery of the GSB of $\mathrm{B}$.
Although, the product of the backward photoisomerization is observed as the negative peak near $(\omega _{1},\omega _{3})=(\omega _{\mathrm{B}},\omega _{\mathrm{A}})$, this is weak because the yield of the backward photoisomerization is small.

For the pulse repetition interval (d) $\tau _{\mathrm{p}}=50~\mathrm{ps}$, the peak intensities near $\omega _{1}=\omega _{\mathrm{A}}$ in Fig.~\ref{fig:2DES}(ii-d) and $\omega =\omega _{\mathrm{A}}$ in Fig.~\ref{fig:2DES}(iii-d) at $t=0$ were further enhanced, because of the thermalization.
While the peak intensity at $t_{2}=0$ is stronger than that in Fig.~\ref{fig:2DES}(ii-c), the intensities of peaks in Figs.~\ref{fig:2DES}(ii-c') and \ref{fig:2DES}(ii-d') are similar.
This is due to the recovery of the GSB from $\mathrm{A}$ proceeded by the thermalization during the longer pulse interval.

As we demonstrated here, using non-linear optical spectra of the stationary rotating process, we can characterize the role of the experimentally controllable repetition time of the driving pulses in the photoisomerization and thermalization processes.
Note that, as shown in Figs.~\ref{fig:2DES}(ii-b') and (ii-c'), while the hindrance of the thermalization $\mathrm{B}\rightarrow \mathrm{A}'$ by the photoexcitation causes an emission peak near $(\omega _{1},\omega _{3})=(\omega _{\mathrm{B}},\omega _{\mathrm{A}})$, the backward photoisomerization ($\mathrm{B}\rightarrow \mathrm{A}$) also causes an absorptive peak near $(\omega _{1},\omega _{3})=(\omega _{\mathrm{B}},\omega _{\mathrm{A}})$.
Because these two peaks may cancel with each other, it is difficult to find a quantitative relation between the rotational process and these optical spectra.

\section{CONCLUSION}
\label{sec:conclusion}
In this paper, we investigated a model for a light-driven molecular motor system, described by a single coordinate with multiple electronic states.
By using the MS-LT-QSE, wavepacket dynamics in the photoisomerization and thermalization processes were simulated.
We analyzed the case that the motor system is driving by repeated laser pulses.
In the case that the timescales of the pulse repetition, photoisomerization, and thermalization are sufficiently separated, the average rotational speed is determined by the timescale of thermalization and the yield of the photoisomerization.
Because the timescale of the thermalization process of the real photo-driven molecular motor is extremely slow (typically $t_{1/2}\sim 1~\si{\micro}\mathrm{s}$--$1~\mathrm{h}$ \cite{pollard2007afm, klok2008jacs}) and well separated from the time scale of the other processes, this condition is fairly realistic.
In this case, we obtain a simple expression for the average rotational speed given in Eq.~\eqref{eq:ipi-approximation}, because the thermalization process is described by a simple rate equation.
In contrast, when the pulse repetition rate is fast and timescales cannot be separated, the detailed dynamics of photoisomerization becomes important.
It was shown that 2D correlation spectra and transient absorption spectra may be helpful to analyze the behavior of the molecular motor system under such high-frequency pulse repetition driving.

Although our analysis in the present paper is limited to a simple one-dimensional model with impulsive excitations, our approach can be extended to study more realistic two-dimensional PESs and non-adiabatic coupling functions.
Because we are solving kinetic equations of motion, we can easily handle arbitrary strength and profile of laser excitations, including a continuous laser irradiation.
To incorporate such time-dependent external fields, we should solve the equations of motion under a time-dependent potential (see for example, Ref.~\onlinecite{kato2013jpcb}).
Analysis of such systems by means of nonlinear spectroscopy is also important, because the peak profiles may be altered by various processes that we did not account for in the present study.
We leave such extensions to future studies, in accordance with progress in experimental and simulational techniques.

\begin{acknowledgments}
  This work was supported by JSPS KAKENHI Grant Number JP26248005.
\end{acknowledgments}

\appendix

\section{DESCRIPTION OF A MULTI-STATE SYSTEM}
\label{sec:hamiltonian-non-adiabatic}
We consider a molecular system expressed by a single effective reaction coordinate $q$ and its conjugate momentum $p$ with multiple electronic adiabatic states, $|\Phi _{a}\rangle $ \cite{mukamel1999book}.
We employ a dimensionless coordinate and its conjugate momentum defined in terms of the actual coordinate and momentum $\bar{q}$ and $\bar{p}$, as $q\equiv \bar{q}\sqrt {m\omega _{0}/\hbar }$ and $p\equiv \bar{p}/\sqrt {m\hbar \omega _{0}}$, where $\omega _{0}$ is the characteristic vibrational frequency of the system and $m$ is the effective mass.
The system Hamiltonian is expressed in the adiabatic representation as
\begin{align}
  \hat{H}(p,q)&\equiv \frac{\hbar \omega _{0}}{2}\hat{p}^{2}+\sum _{ab}|\Phi _{a}\rangle \Bigl(U_{a}(\hat{q})\delta _{ab}+\hat{\Lambda }_{ab}(q)\Bigr)\langle \Phi _{b}|,
  \label{eq:system-Hamiltonian}
\end{align}
where $U_{a}(q)$ is the BO PES of the $a$th adiabatic state.
The non-BO operator $\hat{\Lambda }_{ab}(\hat{q})$ is defined as
\begin{align}
  \hat{\Lambda }_{ab}(q)\equiv -\hbar \omega _{0}\left(id_{ab}(\hat{q})\hat{p}+\frac{1}{2}h_{ab}(\hat{q})\right),
\end{align}
where $d_{ab}(q)$ and $h_{ab}(q)$ are the NAC matrices of the first and second order,
\begin{subequations}
  \begin{align}
    d_{ab}(q)&\equiv \langle \Phi _{a}(q)|\frac{\partial }{\partial q}|\Phi _{b}(q)\rangle \\
    \intertext{and}
    h_{ab}(q)&\equiv \langle \Phi _{a}(q)|\frac{\partial ^{2}}{\partial q^{2}}|\Phi _{b}(q)\rangle ,
  \end{align}
\end{subequations}
respectively \cite{stock2005acp, may2008book}.
These two matrices are related by
\begin{align}
  h_{ab}(q)&=\frac{\partial d_{ab}(q)}{\partial q}+\sum _{c}d_{ac}(q)d_{cb }(q).
  \label{eq:non-adiabatic-coupling-2nd}
\end{align}
The NAC matrix of the first-order, $d_{ab}(q)$, is skew-Hermitian (i.e.~$d_{ba}^{\ast }(q)=-d_{ab}(q)$).
Contrastingly, $h_{ab}(q)$ is neither Hermitian nor skew-Hermitian.
By using $\hat{p}=-i\partial /\partial q$, Eq.~\eqref{eq:non-adiabatic-coupling-2nd} can be rewritten in the matrix form as $\bm{h}(q)=i(\hat{p}\bm{d}(q))+\bm{d}(q)\bm{d}(q)$.
The off-diagonal elements of the non-BO operator, $\hat{\Lambda }_{ab}(\hat{q})$ for $a\neq b$, describes the non-adiabatic transition between the $a$th and $b$th adiabatic states, while the diagonal term, $\Lambda _{aa}(q)$, modulates the $a$th BO PES.
The approximation ignoring the off-diagonal terms in $\hat{\Lambda }_{ab}(\hat{q})$ is referred as the ``Born-Huang (adiabatic) approximation'', and that ignoring whole terms in $\hat{\Lambda }_{ab}(\hat{q})$ is referred as the ``Born-Oppenheimer (adiabatic) approximation'' \cite{azumi1977pp}.
In this paper, we include the whole term $\hat{\Lambda }_{ab}(\hat{q})$ to describe the non-adiabatic transition in the photoisomerization process.

We introduce a scaling parameter, $s$, to redefine the coordinate and momentum as $\theta \equiv q/s$ and $p_{\theta }\equiv ps$ to fit $\theta $  in the region  $\theta =[0,2\pi ]$. Then, by setting $I_{\theta }\equiv s^{2}/\omega _{0}$, Eq.~\eqref{eq:system-Hamiltonian} can be rewritten in terms of $p_{\theta }$ and $\theta $, as presented in Eq.~\eqref{eq:system-Hamiltonian-matrix}.

\section{GREEN'S FUNCTION OF IMPULSIVE PULSE EXCITATION}
\label{sec:pulse-green-function}
In this appendix, we derive Eq.~\eqref{eq:photoexcitation}.
The dipole moment matrix is expressed as $\bm{\mu }=\mu \bm{\sigma }_{x}$, where $\bm{\sigma }_{x}$ is the Pauli matrix.
Because $\bm{\sigma }_{x}^{2}=\bm{1}$, we have
\begin{subequations}
  \begin{align}
    {\bm{\sigma }_{x}^{\times }}^{(2n+1)}\bm{f}&=2^{2n}(\bm{\sigma }_{x}\bm{f}-\bm{f}\bm{\sigma }_{x})\\
    \intertext{and}
    {\bm{\sigma }_{x}^{\times }}^{(2n+2)}\bm{f}&=2^{2n+1}(\bm{f}-\bm{\sigma }_{x}\bm{f}\bm{\sigma }_{x})
  \end{align}
\end{subequations}
for $n\geq 0$.
Therefore, the infinite summation in the matrix exponential in Eq.~\eqref{eq:green-function-p} can be expressed as
\begin{widetext}
  \begin{align}
    \begin{split}
      \mathcal{G}_{\mathrm{p}}\bm{f}&\equiv \exp \left(\frac{i}{\hbar }\bar{E}\Delta \tau \bm{\mu }^{\times }\right)\bm{f}\\
      &=\bm{f}+\sum _{n=0}^{\infty }\frac{1}{(2n+1)!}\left(i\varphi _{\mathrm{p}}\right)^{2n+1}{\bm{\sigma }_{x}^{\times }}^{(2n+1)}+\sum _{n=0}^{\infty }\frac{1}{(2n+2)!}(i\varphi _{\mathrm{p}})^{2n+2}{\bm{\sigma }_{x}^{\times }}^{(2n+2)}\bm{f}\\
      &=\bm{f}+\frac{\sin 2\varphi _{\mathrm{p}}}{2}i(\bm{\sigma }_{x}\bm{f}-\bm{f}\bm{\sigma }_{x})-\frac{1-\cos 2\varphi _{\mathrm{p}}}{2}(\bm{f}-\bm{\sigma }_{x}\bm{f}\bm{\sigma }_{x})\\
      &=(1-\alpha _{\mathrm{E}})\bm{f}+\alpha _{\mathrm{E}}\bm{\sigma }_{x}\bm{f}\bm{\sigma }_{x}+\sin \varphi _{\mathrm{p}}\cos \varphi _{\mathrm{p}}i(\bm{\sigma }_{x}\bm{f}-\bm{f}\bm{\sigma }_{x})\\
      &=(1-\alpha _{\mathrm{E}})
      \begin{pmatrix}
        f_{gg} & f_{ge}\\
        f_{eg} & f_{ee}\\
      \end{pmatrix}
      +\alpha _{\mathrm{E}}
      \begin{pmatrix}
        f_{ee} & f_{eg}\\
        f_{ge} & f_{gg}\\
      \end{pmatrix}
      +i\sin \varphi _{\mathrm{p}}\cos \varphi _{\mathrm{p}}
      \begin{pmatrix}
        f_{eg}-f_{ge} & f_{ee}-f_{gg}\\
        f_{gg}-f_{ee} & f_{ge}-f_{eg}\\
      \end{pmatrix}.
    \end{split}
  \end{align}
\end{widetext}
Here, $\varphi _{\mathrm{p}}\equiv \mu \bar{E}\Delta \tau /\hbar $ and $\alpha _{\mathrm{E}}\equiv \sin ^{2}\varphi _{\mathrm{p}}$.

\section{FITTING FUNCTION FOR NON-ADIABATIC TRANSITION DYNAMICS}
\label{sec:fitting-function}
\begin{figure}
  \centering
  \includegraphics[scale=\SingleColFigScale]{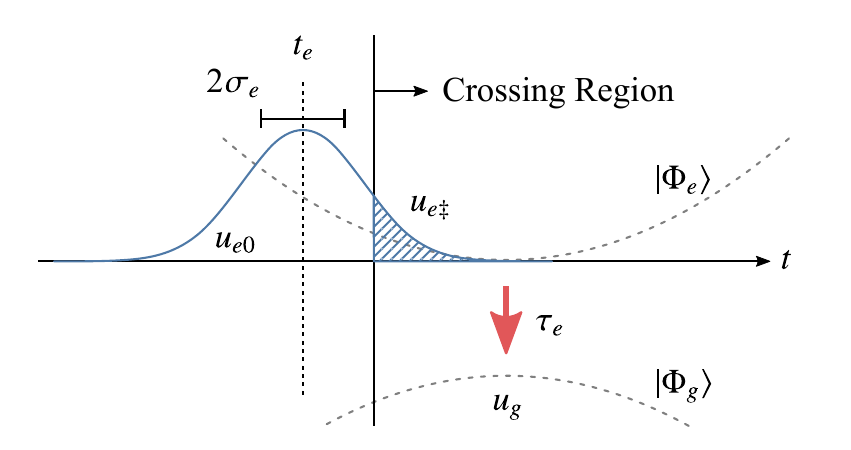}
  \caption{
    The schematic view of the fitting model employed in Eq.~\eqref{eq:fitting-function}.
  }
  \label{fig:fitting-model}
\end{figure}
In this appendix, we derive Eq.~\eqref{eq:fitting-function} as a fitting model.
To simplify the discussion, we assume the following:
The excited wavepacket is expressed as a Gaussian distribution in the coordinate space in the $|\Phi _{e}\rangle $ state.
The wavepacket moves into the crossing region.
The wavepacket is then trapped in the crossing region and de-excites to the $|\Phi _{g}\rangle $ state with the life-time constant $\tau _{e}$.
Because the wavepacket leaves the crossing region in the $|\Phi _{g}\rangle $ state quickly, the transition of the wavepacket from $|\Phi _{e}\rangle $ to $|\Phi _{g}\rangle $ states does not occur.
We denote the population outside the crossing region by $u_{e0}$, that in the crossing region by $u_{e\ddagger }$, and the population in the $|\Phi _{g}\rangle $ state by $u_{g}$.
A schematic illustration of the model is depicted in Fig.~\ref{fig:fitting-model}.
Under the above conditions, the rate equation can be written as
\begin{align}
  \left\{\begin{aligned}
  \frac{du_{e0}}{dt}&=-\frac{1}{\sqrt {\mathstrut 2\pi \sigma _{e}^{2}}}e^{-(t-t_{e})^{2}/2\sigma _{e}^{2}}\\
  \frac{du_{e\ddagger }}{dt}&=-(1/\tau _{e})u_{e\ddagger }+\frac{1}{\sqrt {\mathstrut 2\pi \sigma _{e}^{2}}}e^{-(t-t_{e})^{2}/2\sigma _{e}^{2}}\\
  \frac{du_{g}}{dt}&=+(1/\tau _{e})u_{e\ddagger }.\\
  \end{aligned}\right.
\end{align}
To simplify the formula, we assume that $u_{e0}(-\infty )=1$.
The solution of $u_{e}(t)\equiv u_{e0}(t)+u_{e\ddagger }(t)$ is then expressed as Eq.~\eqref{eq:fitting-function}.

\begin{subequations}
  For $\sigma _{e}^{2}\rightarrow +0$, we have
  \begin{align}
    u_{e}(t)&=\theta (t_{e}-t)+e^{-(t-t_{e})/\tau _{e}}\theta (t-t_{e}),
  \end{align}
  because $\mathrm{erfc}(t/\sqrt {2\sigma _{e}^{2}})/2\rightarrow \theta (-t)$, where $\theta (x)$ is the Heaviside step function.
  This indicates that the time evolution of the population exhibits a discontinuity at $t=t_{e}$.
  For $\tau _{e}\rightarrow +0$, the wavepacket is immediately de-excited after arriving at the crossing region.
  The population dynamics is then approximated by the error function as
  \begin{align}
    u_{e}(t)&=\frac{1}{2}\mathrm{erfc}\left(\frac{t-t_{e}}{\sqrt {\mathstrut 2\sigma _{e}^{2}}}\right).
  \end{align}
\end{subequations}

\section{YIELDS OF THE PHOTOISOMERIZATION}
\label{sec:yields}
In this appendix, we show the numerical calculations to determine the forward ($\mathrm{A}\rightarrow \mathrm{B}$) and backward ($\mathrm{B}\rightarrow \mathrm{A}$) yields of the product state, $y_{\mathrm{P}}^{+}$ and $y_{\mathrm{P}}^{-}$.
For this purpose, we introduce a potential barrier function into the ground BO PES, defined as
\begin{align}
  U_{\mathrm{bar}}(\theta )&\equiv \Delta E_{\mathrm{bar}}\sum _{m=-\infty }^{\infty }e^{-(\theta -\theta _{\mathrm{T}}+\pi m)^{2}/2\sigma _{\mathrm{bar}}^{2}},
\end{align}
to inhibit thermalization during evaluation of the photoisomerization.
Here, $\Delta E_{\mathrm{bar}}$ and $\sigma _{\mathrm{bar}}$ are the height and width of the barrier, respectively.
\begin{subequations}
  We employ two initial distributions, $\bm{f}_{\mathrm{A}}^{\mathrm{loc}}(\theta )$ and $\bm{f}_{\mathrm{B}}^{\mathrm{loc}}(\theta )$, that are localized near $\mathrm{A}$ and $\mathrm{B}$, respectively, as
  \begin{align}
    f_{\mathrm{A},gg}^{\mathrm{loc}}(\theta )&=\left\{
    \begin{matrix}
      e^{-\beta [U_{g}(\theta )+U_{\mathrm{bar}}(\theta )]}/\mathcal{Z}&\text{($0\leq \theta <\theta _{\mathrm{P}}$)}\\
      0&\text{($\theta _{\mathrm{P}}\leq \theta <\pi $)},
    \end{matrix}\right.\\
    f_{\mathrm{B},gg}^{\mathrm{loc}}(\theta )&=\left\{
    \begin{matrix}
      0&\text{($0\leq \theta <\theta _{\mathrm{P}}$)}\\
      e^{-\beta [U_{g}(\theta )+U_{\mathrm{bar}}(\theta )]}/\mathcal{Z}&\text{($\theta _{\mathrm{P}}\leq \theta <\pi $)},
    \end{matrix}\right.
  \end{align}
  and $f_{\mathrm{A}/\mathrm{B},ab}^{\mathrm{loc}}(\theta )=0$ ($a,b=g,e$) otherwise.
\end{subequations}
We then create the excited wavepackets by applying Eq.~\eqref{eq:photoexcitation} with $\alpha _{\mathrm{E}}=1$ to the $f_{\mathrm{A}/\mathrm{B},ab}^{\mathrm{loc}}(\theta )$ and integrate them using the MS-LT-QSE to obtain $f_{\mathrm{A}/\mathrm{B},aa}(\theta , t)$ for sufficiently long time $t=t_{\mathrm{f}}$ for $\mathrm{A}$ and $\mathrm{B}$, respectively.
Using these result, we calculate the population $u_{\mathrm{A}}(t_{\mathrm{f}})$ and $u_{\mathrm{B}}(t_{\mathrm{f}})$, and then the forward yield is evaluated as $y_{\mathrm{P}}^{+}=u_{\mathrm{B}}(t_{\mathrm{f}})$ for the initial distribution $\bm{f}_{\mathrm{A}}^{\mathrm{loc}}(\theta )$, and the backward yield is evaluated as $y_{\mathrm{P}}^{-}=u_{\mathrm{A}}(t_{\mathrm{f}})$ for the initial distribution $\bm{f}_{\mathrm{B}}^{\mathrm{loc}}(\theta )$.

\begin{figure}
  \centering \includegraphics[scale=\SingleColFigScale]{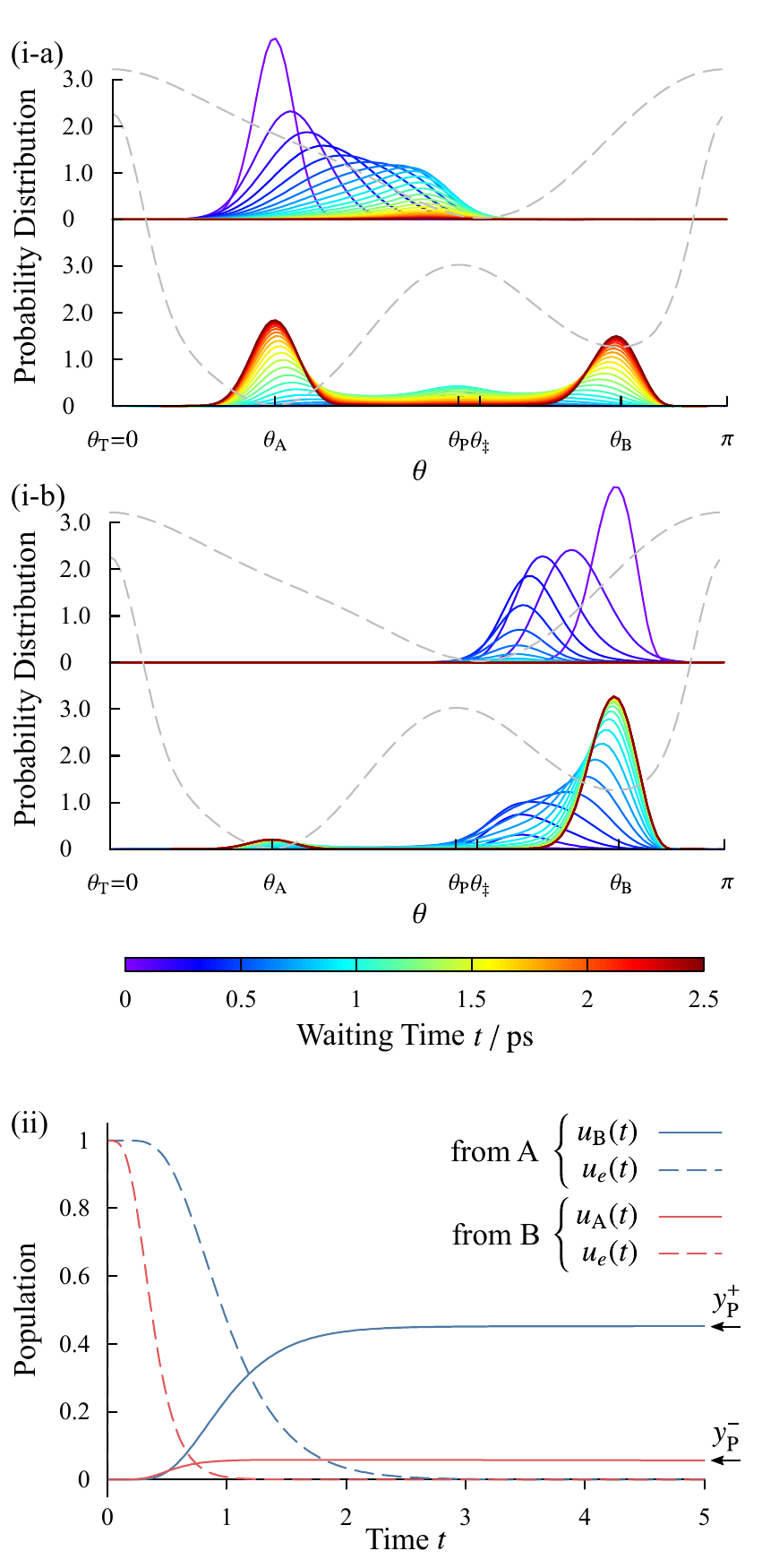}
  \caption{
    (i) Snapshots of wavepackets, $f_{\mathrm{A}/\mathrm{B},aa}(\theta ,t)$ ($a=g,e$), in the adiabatic excited state after the photoexcitation at $t=0$ from the initial state (a) $\bm{f}_{\mathrm{A}}^{\mathrm{loc}}(\theta )$ and (b) $\bm{f}_{\mathrm{B}}^{\mathrm{loc}}(\theta )$.
    The colors of the curves represent the different waiting time after the photoexcitation.
    The gray dotted curves represent the BO PESs, $U_{g}(\theta )+U_{\mathrm{bar}}(\theta )$ and $U_{e}(\theta )$, with an arbitrary unit.
    (ii) Populations as functions of $t$ after the photoexcitation at $t=0$.
    The blue solid curve represents the population near $\mathrm{B}$ from the initial state $\bm{f}_{\mathrm{A}}^{\mathrm{loc}}(\theta )$, whereas the red solid curve represents the population near $\mathrm{A}$ from the initial state $\bm{f}_{\mathrm{A}}^{\mathrm{loc}}(\theta )$.
    The blue and red dashed curve are the populations of the excited state for each case.
  }
  \label{fig:photoisomerization}
\end{figure}
In Fig.~\ref{fig:photoisomerization}, snapshots of wavepackets and population dynamics for the forward and backward photoisomerization processes are displayed.
In the calculations, the parameter values of the barrier were set to $\Delta E_{\mathrm{bar}}=6,000~\mathrm{cm}^{-1}$ and $\sigma _{\mathrm{bar}}=0.15$, and $t_{\mathrm{f}}$ was set to $5~\mathrm{ps}$.
This high and narrow barrier made the numerical calculations difficult.
For this reason, we employed a fine mesh with $N_{q}=128$ with a small timestep $\delta t=0.1\times 10^{-3}~\mathrm{ps}$ in these calculations.
Because the position of the crossing region, $\theta _{\ddagger }$, is closer to $\theta _{\mathrm{B}}$ than to $\theta _{\mathrm{A}}$, the backward photoisomerization is faster than the forward photoisomerization.
Moreover, the excited wavepacket from $\bm{f}_{\mathrm{B}}^{\mathrm{loc}}(\theta )$ is de-excited through the crossing region before arriving the position of the barrier of the photoisomerization, $\theta _{\mathrm{P}}$.
Therefore, the backward yield is very small.
From the numerical calculations, we obtained $y_{\mathrm{P}}^{+}=0.45$ and $y_{\mathrm{P}}^{-}=0.057$.

\section{IPI APPROXIMATION WITH REALISTIC PARAMETERS}
\label{sec:realistic-motor}
In the present study, we evaluated the timescales of the photoisomerization as $t_{e}=0.62~\mathrm{ps}$ and $\tau _{e}=0.41~\mathrm{ps}$, and that of the thermalization processes as $\tau _{\mathrm{T}}=39.5~\mathrm{ps}$.
Thus, they can be analyzed separately to capture the qualitative feature of a photo-driven molecular motor system.
However, the timescale of the thermalization process of the real photo-driven molecular motor is even slower, and typically $t_{1/2}\sim 1~\si{\micro}\mathrm{s}$--$1~\mathrm{h}$ \cite{pollard2007afm, klok2008jacs} (Note that $t_{1/2}=\log (2)\tau _{\mathrm{P}}$).

\begin{figure}
  \centering
\includegraphics[scale=\SingleColFigScale]{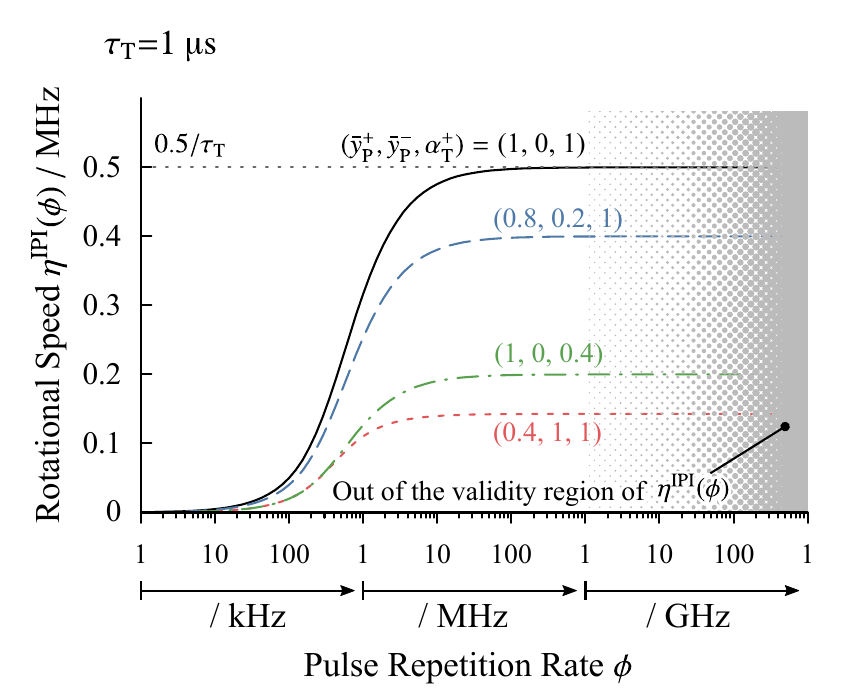}
\caption{
  Average rotational speed in the IPI approximation, Eq.~\eqref{eq:ipi-approximation}, in the case that $\tau _{\mathrm{T}}=1~\si{\micro}\mathrm{s}$ for several values of the parameter sets $(\bar{y}_{\mathrm{P}}^{\mathrm{+}}, \bar{y}_{\mathrm{P}}^{\mathrm{-}}, \alpha _{\mathrm{T}}^{+})$.
}
\label{fig:realistic-motor}
\end{figure}
In Fig.~\ref{fig:realistic-motor}, we plotted the average rotational speed obtained using the IPI approximation, Eq.~\eqref{eq:ipi-approximation}, in the case that $\tau _{\mathrm{T}}=1~\si{\micro}\mathrm{s}$ for several values of the parameter sets $(\bar{y}_{\mathrm{P}}^{\mathrm{+}},\bar{y}_{\mathrm{P}}^{\mathrm{-}},\alpha _{\mathrm{T}}^{+})$.
As the numerical calculations in Sec.~\ref{sec:pulse-repetition} indicate, the IPI approximation breaks down in the region $\phi \gtrsim 100~\mathrm{GHz}$ (i.e.~$1/\phi \lesssim 10~\mathrm{ps}$) in the case that the timescales of the photoisomerization are approximately $1~\mathrm{ps}$, which is approximately the same order observed in the experiment \cite{hall2017jacs}.
However, with a realistic timescale of the thermalization, $\tau _{\mathrm{T}}=1~\si{\micro}\mathrm{s}$, $\eta ^{\mathrm{IPI}}(\phi )$ is enoughly close to it's maximum value, $\eta ^{\mathrm{ID}}$, in the region $\phi \lesssim 1~\mathrm{GHz}$.
Thus, in practice, we can estimate the maximum rotaion speed and the pulse repetition rate to achieve the speed using $\eta ^{\mathrm{IPI}}(\phi )$, in which the timescales of the photoisomerization do not appear.
Note that, the validity of the IPI approximation for the model we employed in this paper was examined in Sec.~\ref{sec:pulse-repetition}.

\let\emph=\textit
\bibliography{ikeda_JCP2018}

\end{document}